\documentclass[lettersize,journal]{IEEEtran}
\usepackage{amsmath,amsfonts, bm}
\usepackage{algorithmic}
\usepackage{algorithm}
\usepackage{array}
\usepackage[caption=false,font=normalsize,labelfont=sf,textfont=sf]{subfig}
\usepackage{textcomp}
\usepackage{stfloats}
\usepackage{url}
\usepackage{verbatim}
\usepackage{graphicx}
\usepackage[nocompress]{cite}
\usepackage[percent]{overpic}
\usepackage{siunitx}
\usepackage{tabularx}
\usepackage{xcolor, soul}
\usepackage[normalem]{ulem}
\begin{document}

\title{A Compact Broadband Purcell Filter for Superconducting Quantum Circuits in a 3D Flip-Chip Architecture}

\author{Zhen Luo,~\IEEEmembership{Graduate Student,~IEEE,}
        Léa Richard,
        Ivan Tsitsilin,
        Anirban Bhattacharjee,
        Christian M.F. Schneider,
        Stefan Filipp and
        Amelie Hagelauer,~\IEEEmembership{Senior Member,~IEEE,} \\
        % <-this % stops a space
%\thanks{This paper was produced by the IEEE Publication Technology Group. They are in Piscataway, NJ.}% <-this % stops a space
%\thanks{Manuscript received April 19, 2021; revised August 16, 2021.}}
\thanks{This work was supported in part by the Federal Ministry of Research, Technology and Space (BMFTR) of Germany as part of Munich Quantum Valley Quantum Computer Demonstrators for Superconducting Qubits (MUNIQC SC) under Project 13N16188, and in part by Munich Quantum Valley (MQV) supported by the Bavarian state government with funds from the Hightech Agenda Bayern Plus, and in part by grants from the EU MSCA Cofund International, Interdisciplinary, and Intersectoral Doctoral Program in Quantum Science and Technologies (QUSTEC) (GrantNr. 847471).}
\thanks{Zhen Luo is with the Technical University of Munich, TUM School of Computation, Information and Technology, Chair of Micro- and Nanosystems Technology, 85748 Garching bei München, Germany (e-mail: zhen.luo@tum.de).}
\thanks{Léa Richard, Ivan Tsitsilin, Anirban Bhattacharjee, Christian Schneider and Stefan Filipp are with the Technical University of Munich, TUM School of Natural Sciences, Department of Physics,
85748 Garching bei München, Germany, and also with Walther-Meißner-Institut, Bayerische Akademie der Wissenschaften, 85748 Garching bei München, Germany.}
\thanks{Amelie Hagelauer is with the Technical University of Munich, TUM School of Computation, Information and Technology, Chair of Micro- and Nanosystems Technology, 85748 Garching bei München, Germany, and also with the Fraunhofer EMFT Institute for Electronic Microsystems and Solid State Technologies, 80686 München, Germany (e-mail: amelie.hagelauer@tum.de).}
% <-this % stops a space}
}

\maketitle

\begin{abstract}
Fast and high-fidelity qubit readout requires strong coupling between the readout resonator and the feedline. However, such coupling unavoidably enhances qubit decay through the Purcell effect. We present a four-pole broadband Purcell filter implemented on a 3D flip-chip platform to overcome this trade-off. The filter provides a flat 1\,GHz passband centered at 7.68\,GHz and achieves more than 45\,dB suppression at typical qubit frequencies. We demonstrate the filter's compatibility with multiplexed readout using a test chip that integrates six floating readout resonators strongly coupled within the passband. The chip is fabricated using a 150\,nm Niobium (Nb) thin-film process and characterized at 20\,mK in a cryogenic measurement setup. We also develop an analytical model that captures the system-level transmission response $S_{21}$, including both the Purcell filter and the in-band coupled readout resonators, directly from their physical geometry, enabling rapid circuit synthesis and design optimization. The proposed design is compact and fabrication-tolerant, making it a practical solution for large-scale superconducting quantum processors.
\end{abstract}


% =======
% FIG. 01
% =======
\begin{figure*}[!t]
    \centering
    \begin{overpic}{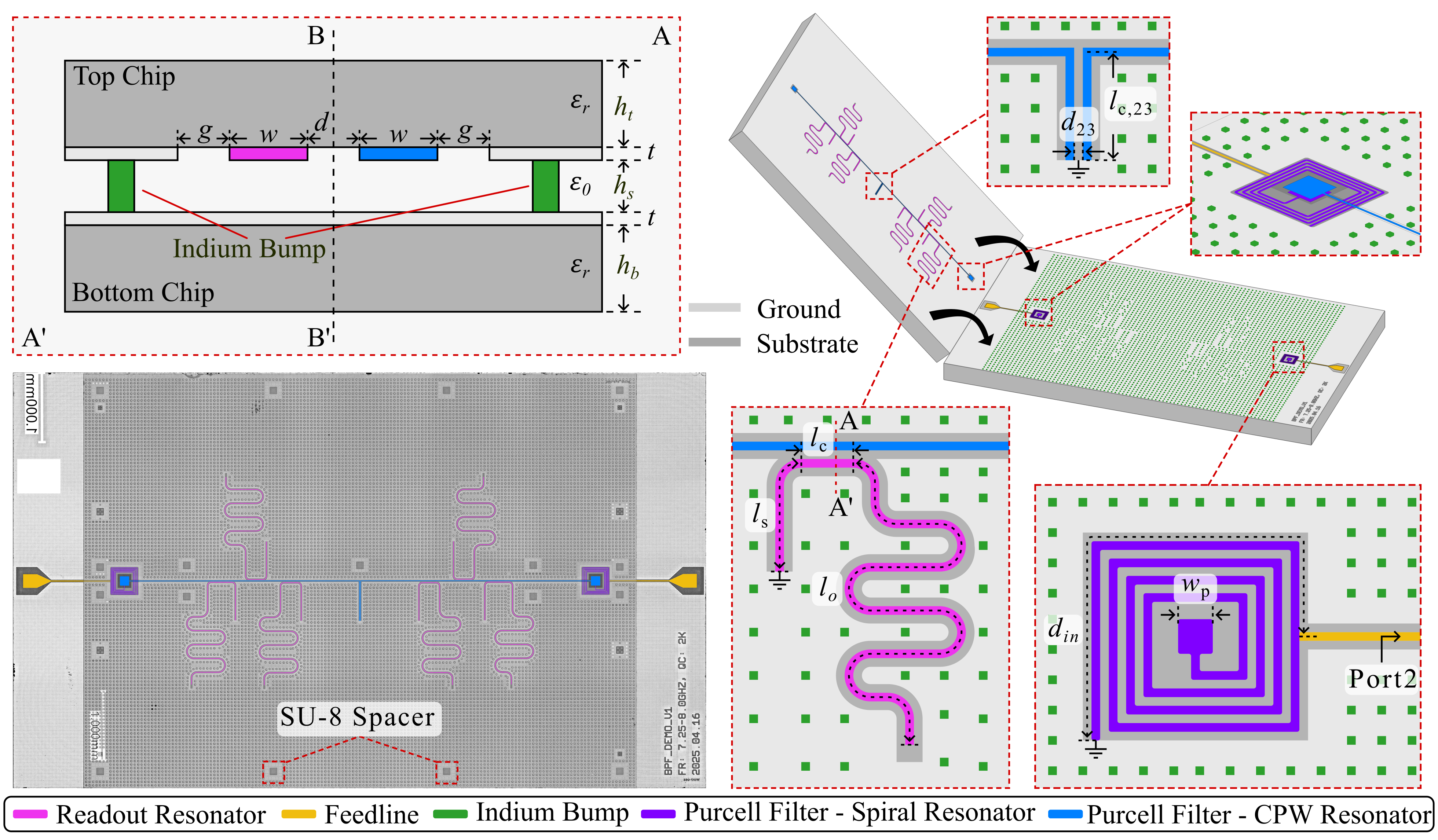}
        \put(1.5, 55){\normalsize (a)}
        \put(1.5, 24.8){\normalsize (b)}
        %\put(38.9, 29.5){\normalsize (c)}
        \put(50, 55){\normalsize (c)}
        \put(65.5, 55){\normalsize (d)}
        \put(83, 51.5){\normalsize (e)}
        \put(51, 31){\normalsize (f)}
        \put(71.5, 26){\normalsize (g)}
    \end{overpic}
    \caption{(a) Cross-sectional view at AA' plane in (f) (not to scale), illustrating the chip stack-up and the coupling section between readout resonator and Purcell filter. The BB' plane denotes the symmetry plane used for conformal mapping analysis. (b) False-colored optical micrograph of the fabricated sample showing the top and bottom chips in an overlapped configuration. (c) Isometric view of the 3D model, unveiling the top chip to show both the metal layers facing each other. The indium bumps are only shown on bottom chip for better visibility. The feedline and spiral resonators are implemented on the bottom chip, while the coplanar waveguide (CPW) line resonator and meandering readout resonators are patterned on the top chip. (d) Coupling section between two CPW line resonators of the Purcell filter. (e) Parallel-plate capacitance between the spiral and CPW line resonators enabling over-the-air coupling. (f) Floating readout resonator coupled to the CPW line resonator within the filter passband. (g) Spiral resonator and tapped-in feedline providing external coupling.}
    \label{fig:main}
\end{figure*}
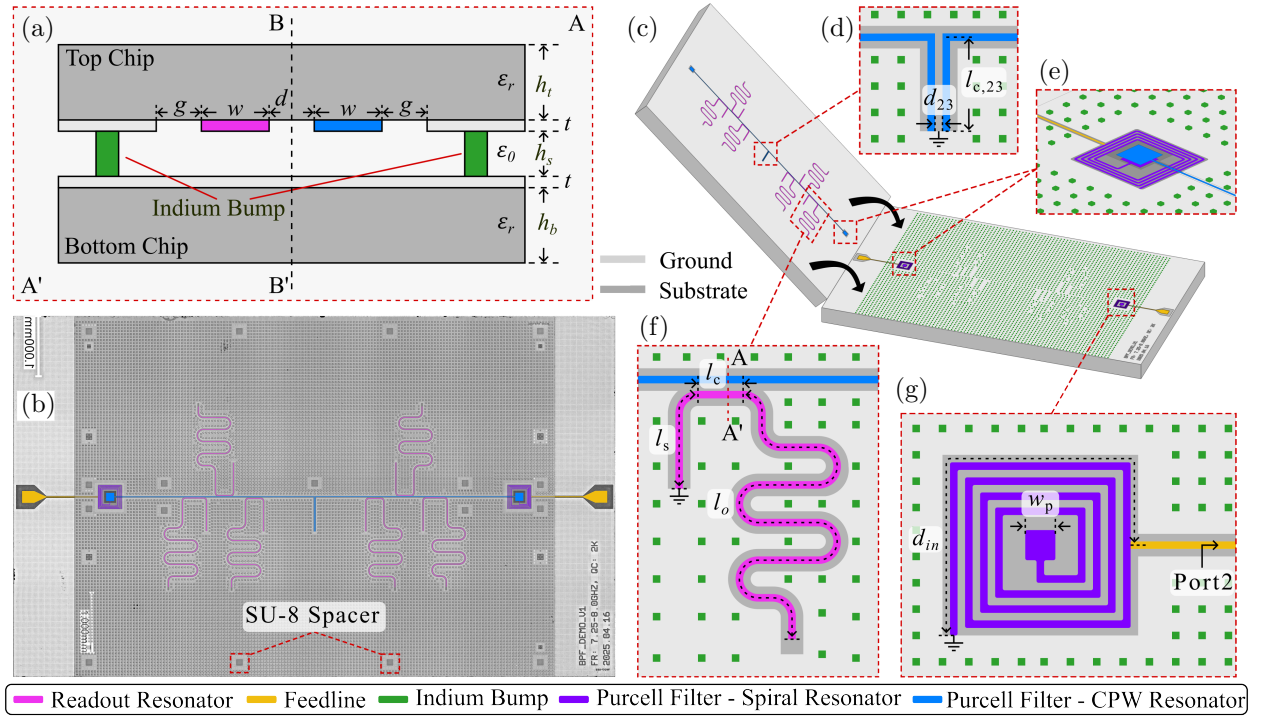

\section{\label{sec:intro} Introduction}
\IEEEPARstart{A}{\lowercase{well-known}} design trade-off in superconducting quantum processors arises from the conflicting requirements of fast readout and long qubit lifetime~\cite{Reed2010, Jeffrey2014, Krantz2019, Blais2021, Walter2017}. Achieving rapid state discrimination requires strong coupling between the readout resonator~\cite{Koch2007} and the feedline, whereas preserving long qubit relaxation time ($T_1$) requires minimizing decay through this same channel~\cite{Purcell1946}. To suppress the unwanted decay while maintaining high readout fidelity, Purcell filters are additively inserted in typical readout circuits~\cite{Houch2008}, \cite{Sete2015}. Such filters are engineered to strongly attenuate emission at the qubit frequency while providing a relative broad transmission band around the resonator frequency, enabling both efficient readout and long qubit lifetime.

In recent implementations of multiplexed readout architectures~\cite{Chen2012}, each readout resonator is often equipped with an individual Purcell filter~\cite{Saxberg2022}, \cite{Heinsoo2018}. These filters are typically realized as half-wavelength ($\lambda/2$) open-ended transmission lines with whose resonance frequencies are nominally identical to those of the associated readout resonators. The Purcell filters are then strongly coupled to a common feedline, enabling frequency-multiplexed readout while maintaining qubit protection. Although this approach has been experimentally demonstrated to provide effective suppression of radiative qubit decay, it requires precise frequency matching between the filter and resonator. The precise frequency targeting and the large footprint of each filter place high demands on design and fabrication, making the realization of large-scale processors increasingly challenging.

An alternative approach integrates a common Purcell filter directly into the feedline, as demonstrated in~\cite{Jeffrey2014}. In that work, a single-pole bandpass filter is realized using a quarter-wavelength ($\lambda/4$) transmission-line resonator, within which four readout resonators are coupled to the filter passband. The suppression at the qubit frequency is around 17\,dB due to the single pole design. The authors achieved readout fidelity exceeding 99.8\,\% within 140\,ns while maintaining qubit relaxation times above 100\,$\mu\text{s}$, allowing fast and high-fidelity measurement through the filter-protected channel. However, the effective passband of this single-pole filter is limited to approximately 200\,MHz. In multiplexed readout architectures, such a narrow transmission window restricts the number of readout resonators that can be accommodated and forces close frequency spacing between qubits, which can affect the frequency discrimination and introduce unwanted crosstalks.

To overcome the bandwidth limitation of single-pole bandpass filters, several groups have implemented multi-pole Purcell filters. In~\cite{Bronn2015}, \cite{Li2024}, stepped-impedance bandpass filters are demonstrated both on- and off-chip, effectively broadening the transmission window and enhancing qubit protection. However, these designs require a relatively large footprint, which becomes problematic for dense qubit integration. Alternatively, in~\cite{Yan2023} an analytical model for a multi-pole broadband Purcell filter is proposed, offering a circuit-level framework for broadband suppression, although no corresponding physical layout is presented. A more recent implementation reported in~\cite{Park2024} employs four coupled resonators to realize a compact broadband Purcell filter with strong suppression near the qubit frequency and a passband of approximately 790\,MHz. The proposed integration strategy, however, is only demonstrated through simulations and lacks experimental verification, and the reported results do not address multiplexed readout operation.

Other approaches have also been explored, such as bandstop filters at the qubit frequency~\cite{Reed2010} and intrinsic Purcell filter configurations integrated directly into the readout resonator structure~\cite{Sunada2022}. However, these methods either exhibit similarly limited passbands or require extensive engineering effort to precisely position the filters. Moreover, all of the aforementioned implementations have so far been demonstrated only on planar platform and have not yet been adapted to the emerging three-dimensional (3D) stacked architectures envisioned for large-scale quantum processors~\cite{Rosenberg2017}. %From these considerations, an effective Purcell filter for next-generation quantum hardware should simultaneously provide a sufficiently wide passband for multiplexed readout, strong suppression at the qubit frequency, a compact footprint, reasonable design complexity, and compatibility with 3D flip-chip integration.

In this work, we present a Purcell filter design based on four coupled resonators that achieves a broad passband of 1\,GHz and 45\,dB suppression near the qubit frequency. To demonstrate its functionality, six floating readout resonators are coupled within the filter’s passband, forming a representative multiplexed readout configuration. An analytical model is developed to guide the filter design efficiently and to reduce the engineering effort required for optimization. The complete system is implemented using a 3D flip-chip architecture, with all components, including the floating readout resonators, integrated on a compact $10\times6\,\text{mm}^2$ chip. Moreover, based on the transmission response $S_{21}$ obtained from the proposed model, the resonance frequency and external quality factor of each readout resonator can be estimated using the method in~\cite{Probst2015}.

% Section 1
\section{\label{sec:method} Circuit Design and Methodology}
The proposed Purcell filter is implemented on a three-dimensional (3D) flip-chip platform, as illustrated in Fig.~\ref{fig:main}(a). %The device consists of two bulk silicon substrates ($\varepsilon_r=11.45$~\cite{Krupka2006}) with thickness $h_t=h_b=500$\,$\mu$m, separated by a fixed spacing of $h_s=10$\,$\mu$m. The metallization layers on both chip are formed by $t=150$\,nm Nb thin-film.To facilitate electrical fan-out of the feedline, the top chip is intentionally made 1\,mm smaller than the bottom chip in lateral dimensions. 
The physical layout of the Purcell filter, shown in the optical micrograph in Fig.~\ref{fig:main}(b), is based on a four-pole bandpass topology comprising four coupled resonators, as illustrated by the corresponding circuit schematic in Fig.~\ref{fig:model}(a). The system consists of three types of resonators serving different roles: spiral resonators acting as the input and output coupling elements, CPW transmission-line resonators forming the inner resonators of the filter, and floating readout resonators coupled within the passband. Each resonator is short-circuited at one end and open-circuited at the other. The feedline is tapped on to the input and output resonators, highlighted in the close-up Fig.~\ref{fig:main}(g), which are implemented as compact spiral structures. The outer terminal of each spiral is connected to ground to form the short-circuited end, while the inner terminal is connected to a square patch of width $w_p^{\text{Spiral}}$, which realizes the open boundary condition. The two inner resonators in the coupling path are realized using unfolded $\lambda/4$ coplanar waveguide (CPW) transmission lines, as depicted in Fig.~\ref{fig:main}(d). The open ends of these resonators are terminated with larger square patches ($w_p^{\text{line}}>w_p^{\text{spiral}}$) %to enhance coupling to the neighboring spiral resonators and 
to mitigate fabrication-induced lateral misalignment, as illustrated in Fig.~\ref{fig:main}(e). Within the coupling path, six floating readout resonators, representing qubit readout channel, are coupled to the two unfolded CPW line resonators via short segments of parallel-coupled lines in Fig.~\ref{fig:main}(f). This configuration emulates a realistic multiplexed readout system.

\subsection{\label{sec:model} Circuit Model and Implementation}
\begin{figure}[b]
    \centering
    \begin{overpic}{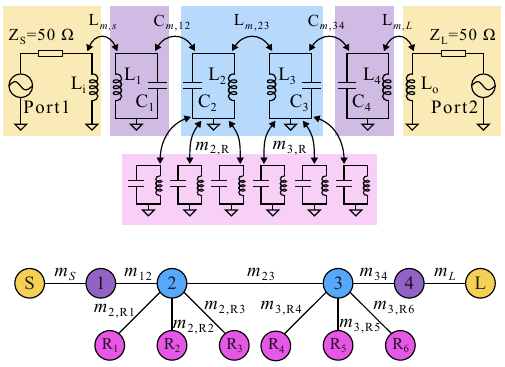}
        \put(-2, 73){\normalsize (a)}
        \put(-2, 23){\normalsize (b)}
    \end{overpic}
    \caption{(a) Schematic of the chip and the coupling scheme between resonators along the signal path. The color coding is consistent with the layout in Fig.~\ref{fig:main}. (b) Coupling topology of the four-pole Purcell filter and the configuration of the in-band coupled readout resonators.}
    \label{fig:model}
\end{figure}

The proposed Purcell filter can be effectively modeled as a network of four coupled parallel LC resonators, as illustrated in Fig.~\ref{fig:model}(a). The feedline is tapped near the virtual ground point of the input and output resonators. Therefore, the coupling between the feedline and these resonators is predominantly inductive, characterized by the mutual inductances $L_{m,S}$ and $L_{m,L}$. These couplings between spiral resonators and CPW line resonators are implemented by parallel plate patch on open-end and denoted as $C_{m,12}$ and $C_{m,34}$. The two inner CPW line resonators are inductively coupled through a short section of parallel coupled lines located near their grounded ends, as shown in Fig.~\ref{fig:main}(d). This interaction is modeled by a mutual inductance $L_{m,23}$. In addition, a set of meandering $\lambda/4$ resonators is coupled to the CPW line resonators within the transmission path.

% =======
% FIG. 02
% =======

Fig.~\ref{fig:model}(b) illustrates the coupling topology of the proposed Purcell filter, together with the in-band coupled readout resonators. The numbered nodes in the equivalent circuit indicate the sequential order of the resonators along the main coupling path. The complete system can be described by a coupling matrix \textit{\textbf{m}} as
\begin{equation}\label{eq:m_matrix}
    \bm{m} = 
    \begin{bmatrix}
    0 & m_S & 0 & 0 & 0 & 0\\
    m_S & m_{11} & m_{12} & 0 & 0 & 0\\
    0 & m_{21} & \bm{m}_{2,R} & m_{23} & 0 & 0 \\
    0 & 0 & m_{32} & \bm{m}_{3,R} & m_{34} & 0 \\
    0 & 0 & 0 & m_{43} & m_{44} & m_L \\
    0 & 0 & 0 & 0 & m_L & 0 \\
    \end{bmatrix},
\end{equation}
where each submatrix $\bm{m}_{i,R} (i=2,3)$ represents the subsystem formed by the \textit{i}-th filter resonator and its associated readout resonators. These submatrices capture the internal couplings within each readout group as well as their interaction with the corresponding filter resonator. In this case, the coupling submatrix $\bm{m}_{i,R}$ can be expressed as
\begin{equation}\label{eq:m_iR}
\bm{m}_{i,R}=
\begin{bmatrix}
m_{ii} & m_{i, j_1} & m_{i, j_2} & m_{i, j_3}\\
m_{i, j_1} & m_{j_1 j_1} & 0 & 0\\
m_{i, j_2} & 0 & m_{j_3 j_3} & 0\\
m_{i, j_3} & 0 & 0 & m_{j_3 j_3}
\end{bmatrix},
\quad i\in\{2,3\}.
\end{equation}
where the index set $(j_1,j_2,j_3)$ corresponds to the three readout resonators coupled to the second filter resonator as $(R1,R2,R3)$ and to the third filter resonator as $(R4,R5,R6)$, respectively.

Once the coupling matrix is determined, the frequency response of the filter can be evaluated in terms of the scattering parameters. The reflection and transmission coefficients are obtained as \cite{Hong2011}
\begin{equation}\label{eq:S21_model}
\begin{aligned}
    S_{11} = 1+2j\left[A\right]_{1,1}^{-1}, \\
    S_{21} = -2j\left[A\right]_{n+2,1}^{-1}, \\
    \bm{A} = \bm{m} + \Omega\bm{U}-j\bm{q},
\end{aligned}
\end{equation}
where \textit{j} is the imaginary unit, $\bm{U}$ denotes a modified $(n+2)\times(n+2)$ identity matrix with boundary elements $[U]_{11}=[U]_{n+2,n+2}=0$, and $\bm{q}$ is an $(n+2)\times(n+2)$ diagonal matrix with all entries zeros, except for $[q]_{11}=[q]_{n+2,n+2}=1$. Assuming the filter is designed to have a passband $f_l - f_h$, where $f_l$ and $f_h$ indicate the passband-edge frequencies. The normalized frequency variable $\Omega$ is introduced to map the actual frequency response onto the low-pass prototype domain as
\begin{equation}
    \Omega=\frac{1}{FBW}\left(\frac{f}{f_0} - \frac{f_0}{f} \right),  
\end{equation}
with $f_0$ and $FBW$ being the center frequency and fractional bandwidth of bandpass filter, respectively, given by
\begin{equation}
\begin{aligned}
    f_0 &= \sqrt{f_h \cdot f_l},\quad FBW = \frac{f_h - f_l}{f_0}.
\end{aligned}
\end{equation}
\begin{comment}
    The entries in the coupling matrix in ~(\ref{eq:m_matrix}) can be categorized into three groups: the self-coupling coefficients $m_{ii}$ on diagonal, which determine the individual resonant frequencies; the mutual coupling coefficients $m_{ij}$ off the diagonal, which define the coupling strength between adjacent resonators; and the external coupling coefficients $m_S$ and $m_L$, which describe the interaction between the filter and the feedline at the input and output ports, respectively. In the following subsections, we detail the procedures used to extract each of these parameters.
\end{comment}

The entries in the coupling matrix in~(\ref{eq:m_matrix}) can be categorized into three groups: the self coupling coefficients $m_{ii}$ on the diagonal; the off-diagonal mutual coupling coefficients $m_{ij}, i\neq j$; and the external coupling coefficients $m_S$ and $m_L$. %In the following subsections, we describe the procedures used to extract each of these parameters.

\subsection{\label{sec:M_ii} Extracting Self Coupling Coefficient}
The diagonal elements $m_{ii}$ of the coupling matrix in~(\ref{eq:m_matrix}) determine the uncoupled resonance frequencies $f_{0i}$ of the individual resonators in the absence of other resonators in the circuit. Each resonator’s bare frequency is governed primarily by its electrical length and cross-sectional geometry. For a $\lambda/4$ transmission-line resonator, the resonance frequency can be expressed as
\begin{equation}\label{eq:bare_f}
f_{r} = \frac{1}{4 l_{t} \sqrt{L_l^g C_l^g}},
\end{equation}
where $l_t$ is the physical length of the resonator and $L_l^g$ and $C_l^g$ are the geometrical inductance and capacitance per unit length, respectively. These parameters can be determined using the conformal mapping technique described in Appendix~\ref{app:TGCPW}. 

The square capacitive patch, with capacitance of $C_p$, located at the open end of the filter resonator introduces an additional capacitive loading that can be approximated as an equivalent extension of the resonator’s physical length, such that
\begin{equation}\label{eq:l_eff}
    l_{eff}=l_t+\Delta{l},\quad \Delta{l} = \frac{C_p}{C_l^g}.
\end{equation}

It should be noted that the spiral resonators differ from standard CPW structures, as they lack a parallel ground plane on one side. Consequently, the conformal mapping approach described in Appendix~\ref{app:TGCPW} is not directly applicable. For these resonators, EM simulations are employed to obtain accurate estimates of their fundamental resonance frequencies and corresponding self-coupling coefficients.

The normalized self coupling coefficient corresponding to each resonator is defined as
\begin{equation}\label{eq:m_ii}
    m_{ii} = 2\cdot\frac{f_{0i} - f_0}{FBW \cdot f_0}.
\end{equation}

\subsection{\label{sec:M_ij} Extracting Mutual Coupling Coefficient}
\begin{figure}[b]
    \includegraphics{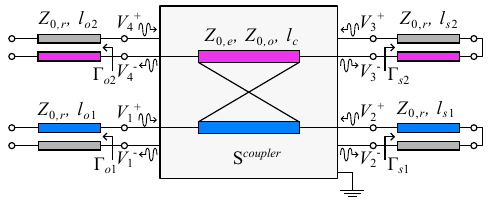}% Here is how to import EPS art
    \caption{General distributed-element model of two coupled $\lambda/4$ resonators, with short- and open-circuit boundary conditions applied to the respective ports.}
    \label{fig:coupled_resonator}
\end{figure}

The mutual coupling coefficient $m_{ij}$ quantifies the electromagnetic coupling strength between adjacent resonators in Fig.~\ref{fig:model}(b). A general equivalent circuit for two coupled $\lambda/4$ resonators is illustrated in Fig.~\ref{fig:coupled_resonator}.  Each resonator consists of three sections: a short-circuited section of length $l_{s,i}$, a coupled section of length $l_{c}$, and an open-ended section of length $l_{o,i}$. The total electrical length of the \textit{i}-th resonator is therefore $l_{t,i} = l_{s,i}+l_{c}+l_{o,i}$.

To analyze the coupling section highlighted by the ivory box in Fig.~\ref{fig:coupled_resonator}, the structure is modeled as a four-port network. This coupled-line section can be characterized using even–odd mode analysis, as detailed in Appendix~\ref{app:coupler}, and represented in terms of its impedance matrix $\mathbf{Z^{coupler}}$ given in~(\ref{eq:coupler_z_matrix}). The matrix elements in $\mathbf{Z^{coupler}}$ are determined by the cross-sectional geometry of the coupled lines and can be evaluated analytically using the conformal mapping technique described in Appendix~\ref{app:Coupled_TGCPW}.

\begin{comment}
    The mutual coupling coefficient $m_{ij}$ quantifies the electromagnetic coupling strength between adjacent resonators in Fig.~\ref{fig:model}(b). A general equivalent circuit for two coupled $\lambda/4$ resonators is illustrated in Fig.~\ref{fig:coupled_resonator}.  Each resonator consists of three sections: a short-circuited section of length $l_{s,i}$, a coupled section of length $l_{c}$, and an open-ended section of length $l_{o,i}$. The total electrical length of the \textit{i}-th resonator is therefore $l_{t,i} = l_{s,i}+l_{c}+l_{o,i}$.

    To analyze the coupling section highlighted by the ivory box in Fig.~\ref{fig:coupled_resonator}, the structure is modeled as a four-port network. This coupled-line section can be characterized using even–odd mode analysis, as detailed in Appendix~\ref{app:coupler}, and represented in terms of its impedance matrix $\mathbf{Z^{coupler}}$ given in~(\ref{eq:coupler_z_matrix}). The matrix elements in $\mathbf{Z^{coupler}}$ are determined by the cross-sectional geometry of the coupled lines and can be evaluated analytically using the conformal mapping technique described in Appendix~\ref{app:Coupled_TGCPW}.
\end{comment}

 % =======
% FIG. 03
% =======

The corresponding scattering matrix of the coupling section, $\mathbf{S}^{\text{coupler}}$, can then be obtained from the impedance matrix through the standard network transformation
\begin{equation}\label{eq:s_matrix}
\mathbf{S}^{\text{coupler}} = \left(\mathbf{Y}_0 \cdot \mathbf{Z}^{\text{coupler}} - \mathbf{Z}_0\right) \cdot \left(\mathbf{Y}_0 \cdot \mathbf{Z}^{\text{coupler}} + \mathbf{Z}_0 \right)^{-1},
\end{equation}
where $\mathbf{Z_0}$ is a diagonal matrix having the square root of the characteristic impedance at each port $\mathbf{Z_0}=\mathrm{diag}\left[\sqrt{Z_{01}},...,\sqrt{Z_{04}}\right]$ and $\mathbf{Y_0}=\mathbf{Z_0}^{-1}$.

By definition, the scattering parameters relate the incident and reflected voltage waves at each port as
\begin{equation}\label{eq:S_definition}
    V_{i}^-=\sum_j S_{ij}^{\text{coupler}}V_i^+.
\end{equation}
where $V_i^+$ and $V_i^-$ denote the incident and reflected voltage wave at \textit{i}-th port.

When each port of the coupling section is terminated in either a short or an open boundary condition, the relationship between the incident and reflected voltage waves can be written in terms of the corresponding reflection coefficients as
 \begin{equation}\label{eq:S_gamma}
\mathbf{V^+}=\space\mathbf{\Gamma}\cdot \mathbf{V^-},
\end{equation}
where $\mathbf{\Gamma}=\mathrm{diag}\left[\Gamma_{o1},\Gamma_{s1}, \Gamma_{s2}, \Gamma_{o2}\right]$ is a diagonal matrix collecting the port reflection coefficients. The reflection coefficients $\Gamma_{s,i}$ and $\Gamma_{o,i}$, seen looking toward the short- and open-end sections of resonator \textit{i}, are determined by the corresponding section lengths, $l_{s,i}$ and $l_{o,i}$:
\begin{align}
    \Gamma_{s,i} &= \frac{jZ_{0,r}\tan(\beta l_{s,i}) - Z_{0}}{jZ_{0,r}\tan(\beta l_{s,i}) + Z_{0}},\label{gamma_s} \\
    \Gamma_{o,i} &= \frac{-jZ_{0,r}\cot(\beta l_{o,i}) - Z_{0}}{-jZ_{0,r}\cot(\beta l_{o,i}) + Z_{0}}, \label{gamma_o}.
\end{align}
Here, $Z_{0,r}$ is the characteristic impedance of the resonator transmission line, which can be obtained using~(\ref{eq:Z0_cpw}) via the conformal-mapping method in Appendix~\ref{app:TGCPW}, and $Z_0$ is the port reference impedance used in~(\ref{eq:s_matrix}). In most practical cases, all transmission lines in the system are designed with equal characteristic impedance, i.e., $Z_{0,r}=Z_0=50\,\Omega$. Under this condition, the reflection coefficients in (\ref{gamma_s}) and (\ref{gamma_o}) are simplified to $\Gamma_s=-1e^{-j2\beta l_s}$ and $\Gamma_o=1e^{-j2\beta l_o}$.

Substituting the scattering relation of~(\ref{eq:S_definition}) into the termination condition of~(\ref{eq:S_gamma}) yields the characteristic equation for the coupled system. The dressed eigenfrequencies of the two-resonator pair are obtained as the roots of
 \begin{equation}\label{eq:dressed_f}
\mathrm{det}(\mathbf{I} - \mathbf{\Gamma} \cdot \mathbf{S}^{\text{coupler}})=0.
\end{equation}

Let $f_1$ and $f_2$ denote the two solutions of~(\ref{eq:dressed_f}) in the vicinity of the bare resonance $f_{0i}$ from~(\ref{eq:bare_f}). The mutual coupling coefficient is then extracted by \cite{Hong2011}
 \begin{equation}\label{eq:m_ij}
     M_{ij} = \pm \frac{1}{2}\left(\frac{f_{02}}{f_{01}} + \frac{f_{01}}{f_{02}} \right) \sqrt{\left(\frac{f_2^2-f_1^2}{f_2^2+f_1^2} \right)^2-\left(\frac{f_{02}^2-f_{01}^2}{f_{02}^2+f_{01}^2} \right)^2},
 \end{equation}
and the corresponding normalized value for the entries in~(\ref{eq:m_matrix}) is $m_{ij}= M_{ij}/FBW$.

In practice, the coupling coefficient between filter resonator $i=2$ and its associated readout resonator $j \in \{R1, R2, R3\}$ (or $j \in\{R4, R5, R6\}$ for $i=3$) in~(\ref{eq:m_iR}) is evaluated using the general model of two coupled $\lambda/4$ resonators developed above. The coupling coefficient between the two CPW line resonators, $m_{23}$ in Fig.~\ref{fig:model}(b), is also obtained from the same model by taking the limiting case in which the short sections vanish, $l_{s1}=l_{s2}=0$. As shown in Fig.~\ref{fig:main}(d), the coupling section starts directly at the short-end of the $\lambda/4$ resonator. Note that the effective length of open section is compensated using~(\ref{eq:l_eff}) to account the effect of the self-capacitance of the capacitive patch at open-end. 

\begin{figure}[h]
    \includegraphics{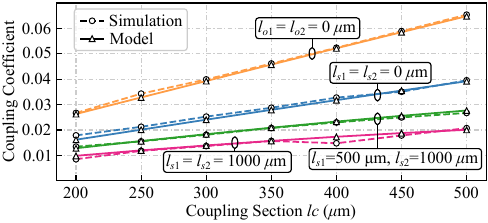}% Here is how to import EPS art
    \caption{ Extracted coupling coefficients of two coupled $\lambda/4$ resonators with different geometric parameters, obtained from the proposed analytical model (solid lines) and Ansys HFSS eigenmode simulations (dash lines).}
    \label{fig:Model_vs_sim}
\end{figure}

To assess the validity and robustness of the model, we benchmark it against Ansys HFSS eigenmode simulations. The test structure comprises two coupled $\lambda/4$ resonators with total lengths $l_{t1}=4340\,\mu\text{m}$ and $l_{t2}=4400\,\mu\text{m}$, and a fixed coupled-line spacing $d=10\,\mu\text{m}$. We sweep the coupling-section length $l_c$ over a representative range and evaluate four boundary configurations: general cases with $l_{o,i} \not =0\,\mu\text{m}$ and $l_{s,i} \not =0\,\mu\text{m}$, as well as two limiting cases with $l_{s1}=l_{s2}=0\,\mu\text{m}$ and $l_{o1}=l_{o2}=0\,\mu\text{m}$. For each configuration, we extract the bare and dressed resonance frequencies either from the model or from HFSS eigenmode simulations. Because HFSS provides eigenfrequencies rather than coupling coefficients, the simulation traces in Fig.~\ref{fig:Model_vs_sim} are obtained by post-processing those frequencies with~(\ref{eq:m_ij}), using the same definition as in the model to ensure a consistent comparison. As shown in Fig.~\ref{fig:Model_vs_sim}, the value closely tracks the HFSS simulation results across the entire sweep, indicating that the proposed formulation is robust and effective. The data also highlight that the coupling strength is governed not only by the coupling-section length $l_c$ but also by the relative position of the coupling-section along the resonators.

 % =======
% FIG. 04
% =======

\begin{figure}[!h]
    \includegraphics{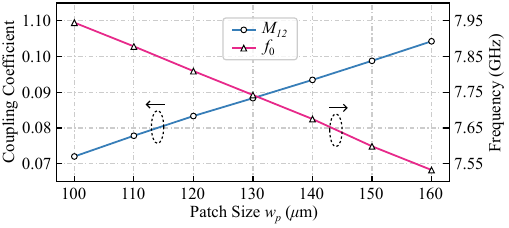}%
    \caption{Extracted coupling coefficients $M_{12}$ between spiral and CPW line resonator, and the corresponding bare frequency of the CPW line resonator, obtained from Ansys HFSS eigenmode simulations. The patch size of the CPW line resonator $w_{p}^{\text{line}}$ is swept, while the patch size of the spiral resonator is fixed at $w_{p}^{\text{spiral}} = 110\,\mu\text{m}$.}
    \label{fig:M12_vs_wp}
\end{figure}

Unlike the coupled-line couplings discussed above, the couplings between the spiral and CPW line resonators, $m_{12}$ and $m_{34}$, are realized by a parallel-plate capacitance placed at their open ends. Therefore, the proposed model does not apply. In this case the capacitive coupling is governed by the mutual capacitance $C_m$ between the facing patches, which is expressed as
\begin{equation}\label{m_12}
    m_{12} = \frac{C_m}{\sqrt{C1 \cdot C2}},
\end{equation}
where $C_1$ and $C_2$ are the equivalent capacitance of first and second resonator in Fig.~\ref{fig:model}(a), respectively. Since the chip spacing $h_s=10\,\mu\text{m}$ is fixed by the fabrication process, the most effective tuning parameter for $m_{12}$ (and $m_{34}$) is the effective area of the parallel-plate capacitor that determines $C_{m}$. Although an ideal parallel-plate capacitor admits a closed-form expression, in our design the top patch is deliberately enlarged relative to the bottom patch to mitigate sensitivity to lateral misalignment, as shown in Fig.~\ref{fig:main}(e). The resulting fringing fields invalidate simple analytical formulas. In practice, we determine $m_{12}$ (and $m_{34}$) by performing eigenmode simulations in Ansys HFSS to extract both the bare and dressed frequencies, and subsequently use~(\ref{eq:m_ij}) to calculate the coupling coefficient. As shown in Fig.~\ref{fig:M12_vs_wp}, varying the size of the top patch effectively tunes the coupling coefficient, while also slightly affecting the bare frequency.

% =======
% FIG. 05
% =======

\subsection{\label{sec:ms_ml} Extracting Input and Output Coupling}
The external coupling coefficients, $m_S$ and $m_L$, characterize the interactions between the filter and the feedline at the input and output ports, respectively. In this design, both external couplings are implemented using a 50\,$\Omega$ tapped feedline, as shown in Fig.~\ref{fig:main}(g). The strength of each coupling can be quantified through the corresponding external quality factor, defined as
\begin{equation}\label{eq:qe}
Q_e = \frac{\omega_0 \cdot \tau_{S_{11}} (\omega_0)}{4},
\end{equation}
where $\omega_0$ is the resonant angular frequency and $\tau_{S_{11}}(\omega_0)$ denotes the group delay of $S_{11}$ at resonance. Once $Q_e$ is obtained, the normalized coupling coefficient $m_S$ (and $m_L$) in~(\ref{eq:m_matrix}) can be extracted as
\begin{equation}\label{eq:ms}
m_s = \sqrt{\frac{1}{Q_e \cdot FBW}}.
\end{equation}

\begin{figure}[!t]
    \includegraphics{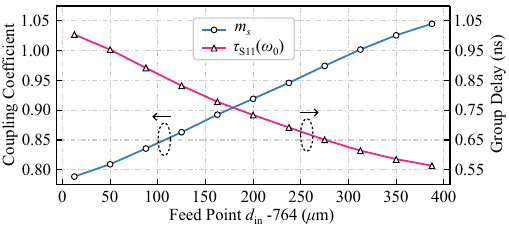}%
    \caption{Group delay of $S_{11}$ obtained from Ansys HFSS driven-modal simulations and the extracted normalized coupling coefficients $m_S$. The tick labels are referenced to the feed point after the first two turns of the spiral resonator. An offset of 764\,$\mu$m accounts for the length of the path from the short-circuit end along the spiral's outer diameter.}
    \label{fig:coupling_vs_din}
\end{figure}

Fig~\ref{fig:coupling_vs_din} presents the simulated group delay of $S_{11}$ at resonance for various positions of tapped-line $d_{in}$, together with the corresponding coupling coefficients extracted using~(\ref{eq:ms}). The parameter $d_{in}$ denotes the distance between the feed point and the short end of spiral resonator. When $d_{in}=0\,\mu\text{m}$, the feedline is directly connected to the short circuit, resulting in a strong impedance mismatch and, consequently, a pronounced reflection toward the source. In this configuration, the external coupling is minimal. Increasing $d_{in}$ moves the feed point away from the shorted end, thereby enhancing the coupling strength, as clearly observed in Fig.~\ref{fig:coupling_vs_din}.

% =======
% FIG. 06
% =======

To this end, all coupling parameters in~(\ref{eq:m_matrix}) are determined as functions of the relevant geometric dimensions. With the coupling matrix established, the overall transmission response $S_{21}$ of the Purcell filter and the multiplexed readout resonators within the passband can be computed using~(\ref{eq:S21_model}). 

\section{\label{sec:Experiments} Experimental Characterization}
% =======
% Tabel 1
% =======
{
\begin{table}[b]
\setlength\tabcolsep{12pt}
\caption{Geometric parameters of the Purcell filter.}
\label{tab:geo_filter}
\centering
\normalsize
\begin{tabular}{lcc}
\hline
Parameter & & Value \\
\hline
Spiral total length, $l_t^{\mathrm{spiral}}$ ($\mu\text{m}$) & & $4904$ \\
Line total length, $l_t^{\mathrm{line}}$ ($\mu\text{m}$) &  & $3900$ \\
CPW line width, $w$ ($\mu\text{m}$) &  & $10$ \\
CPW line gap $g$ ($\mu\text{m}$) &  & $9$ \\
Spiral outer diameter, $d_o^{\mathrm{spiral}}$ ($\mu\text{m}$) & & $375$ \\
Bottom patch width $w_{p}^{\mathrm{spiral}}$ ($\mu\text{m}$) &  & $110$ \\
Top patch width $w_{p}^{\mathrm{line}}$ ($\mu\text{m}$) &  & $135$ \\
Coupling-section length $lc_{c,23}$\textsuperscript{a} ($\mu\text{m}$) &  & $575$ \\
Coupling-section spacing $d_{23}$\textsuperscript{b} ($\mu\text{m}$) &  & $6$ \\
\hline
\end{tabular}
\par\vspace{2pt}
{\begin{minipage}{0.9\columnwidth}
\raggedright 
\footnotesize
\textsuperscript{a}The length of the coupling section of the edge-coupled CPW line resonator 2 and 3.\par
\textsuperscript{b}The edge-to-edge spacing between the inner strips of the CPW line resonator 2 and 3 within the coupling section.\par
\end{minipage}}
\end{table}
}
We design and fabricate a test chip on the 3D flip-chip platform to characterize the experimental performance of the proposed Purcell filter for multiplexed readout. The filter is designed to provide a flat passband from $f_l=7.20\,\text{GHz}$ to $f_h=8.20\,\text{GHz}$, corresponding to a center frequency $f_0=7.68\,\text{GHz}$ and a fractional bandwidth $FBW=13.2\,\%$. 

Table~\ref{tab:geo_filter} summarizes the geometric parameters of the spiral and CPW line resonators used in the Purcell filter. Bare resonance frequencies obtained from Ansys HFSS eigenmode simulations are $f_0^\text{spiral}=7.68\,\text{GHz}$ and $f_0^\text{line}=7.70\,\text{GHz}$. The corresponding normalized coupling parameters for this design are extracted using the procedures described in Section~\ref{sec:M_ii} and Section~\ref{sec:M_ij} as
\begin{gather*}
    m_{11} = m_{44} = -0.0075, \quad m_{22} = m_{33} = 0.0325,\\
    m_{12} = m_{34} = 0.6608, \quad m_{23} = 0.4861.
\end{gather*}

The tapped-line feed point for both the input and output resonators is located at $d_{in}=951.5\,\mu\text{m}$. At the resonance frequency of the spiral resonator $\omega_0^{\text{spiral}}$, the group delay of the reflection coefficient is $\tau_{S_{11}}(\omega_0^{\text{spiral}})=0.733\,\text{ns}$, obtained from Ansys HFSS driven-modal simulation. Using~(\ref{eq:ms}) this corresponds to normalized external coupling coefficients $m_S=m_L=0.93$ for the source and load ports.

Six floating readout resonators are integrated within the passband of the Purcell filter with around 100\,MHz frequency spacing to each other on spectrum. The geometric parameters are given in Table~\ref{tab:gep_resonators}. The bare frequencies and the corresponding self coupling coefficients are obtained from~(\ref{eq:bare_f}) and~(\ref{eq:m_ii}), respectively. Each resonator is coupled to one of the two inner CPW line resonators of the Purcell filter and the dressed frequencies together with mutual coupling coefficients are determined using~(\ref{eq:dressed_f}) and~(\ref{eq:m_ij}). 

With all coupling parameters determined, the calculated transmission response $S_{21}^{\text{cal.}}$ is obtained from~(\ref{eq:S21_model}), and a full-wave simulation of the device using Ansys HFSS (driven-modal) yields $S_{21}^{\text{sim.}}$ for reference. These responses are compared with the measurement data in Section~\ref{sec:Results}. 

\begin{comment}
    The geometric parameters of the multiplexed readout resonators coupled within the passband of the Purcell filter are listed in Table~\ref{tab:gep_resonators}. The bare frequencies follow from~(\ref{eq:bare_f}), and the corresponding self-coupling coefficients are obtained using~(\ref{eq:m_ii}) as
\begin{gather*}
    m_{R1, R1} = 0.5177, \quad m_{R2, R2} = 0.3470,\\
    m_{R3, R3} = 0.1724, \quad m_{R4, R4} = -0.0064,\\
    m_{R5, R5} = -0.1894, \quad m_{R6, R6} = -0.3769.
\end{gather*}

Applying the general model of coupled $\lambda/4$ resonators in Section~\ref{sec:M_ij}, we can determine the dressed frequencies and subsequently extract the mutual coupling coefficients in~(\ref{eq:m_iR}) as
\begin{gather*}
    m_{2, R1} = 0.0805, \quad m_{2, R2} = 0.0805, \quad m_{2, R3} = 0.0691, \\
    m_{3, R4} = 0.0613, \quad m_{3, R5} = 0.0559, \quad m_{3, R6} = 0.0788.
\end{gather*}

\end{comment}

\begin{comment}
    After extracting all the coupling parameters, we compute the transmission spectrum $S_{21, \text{cal.}}$ using~(\ref{eq:S21_model}). For comparison, we simulate the device with Ansys HFSS (driven-modal) to obtain a simulated $S_{21, \text{sim.}}$. The characteristic parameters of the floating readout resonators, $f_0$ and $Q_e$, extracted from the calculated and simulated $S_{21}$ are summarized in Table~\ref{tab:gep_resonators}.
\end{comment}

\subsection{\label{sec:fab} Device Implementation}
The chip stack-up used for the chip design and simulation is shown in Fig.~\ref{fig:main}(a) and consists of two bulk silicon substrates ($\varepsilon_r=11.45$~\cite{Krupka2006}) with thickness $h_t=h_b=500\,\mu\text{m}$, separated by a chip spacing $h_s=10\,\mu\text{m}$ defined by SU-8 polymer spacers. Both top and bottom chips are patterned with 150\,nm thick Nb films for metal layers. The two chips are assembled using a flip-chip bonding process adapted from~\cite{Norris2024}. Superconducting indium bumps are deposited on both chips to provide galvanic inter-chip connections. The bottom chip size is $10\times6\,\text{mm}^2$, while the top chip is intentionally made 1\,mm smaller than the bottom chip in lateral dimensions to facilitate electrical fan-out of the feedline. An optical micrograph of the assembled device is shown in Fig.~\ref{fig:main}(b). The fabrication process is identical to that used for qubit-integrated 3D flip-chip devices, thereby inherently incorporating fabrication tolerance encountered in practical superconducting quantum processors.

\subsection{\label{sec:Setup} Measurement Setup}
% =======
% FIG. 7
% =======
\begin{figure}[!b]
    \centering
    \begin{overpic}{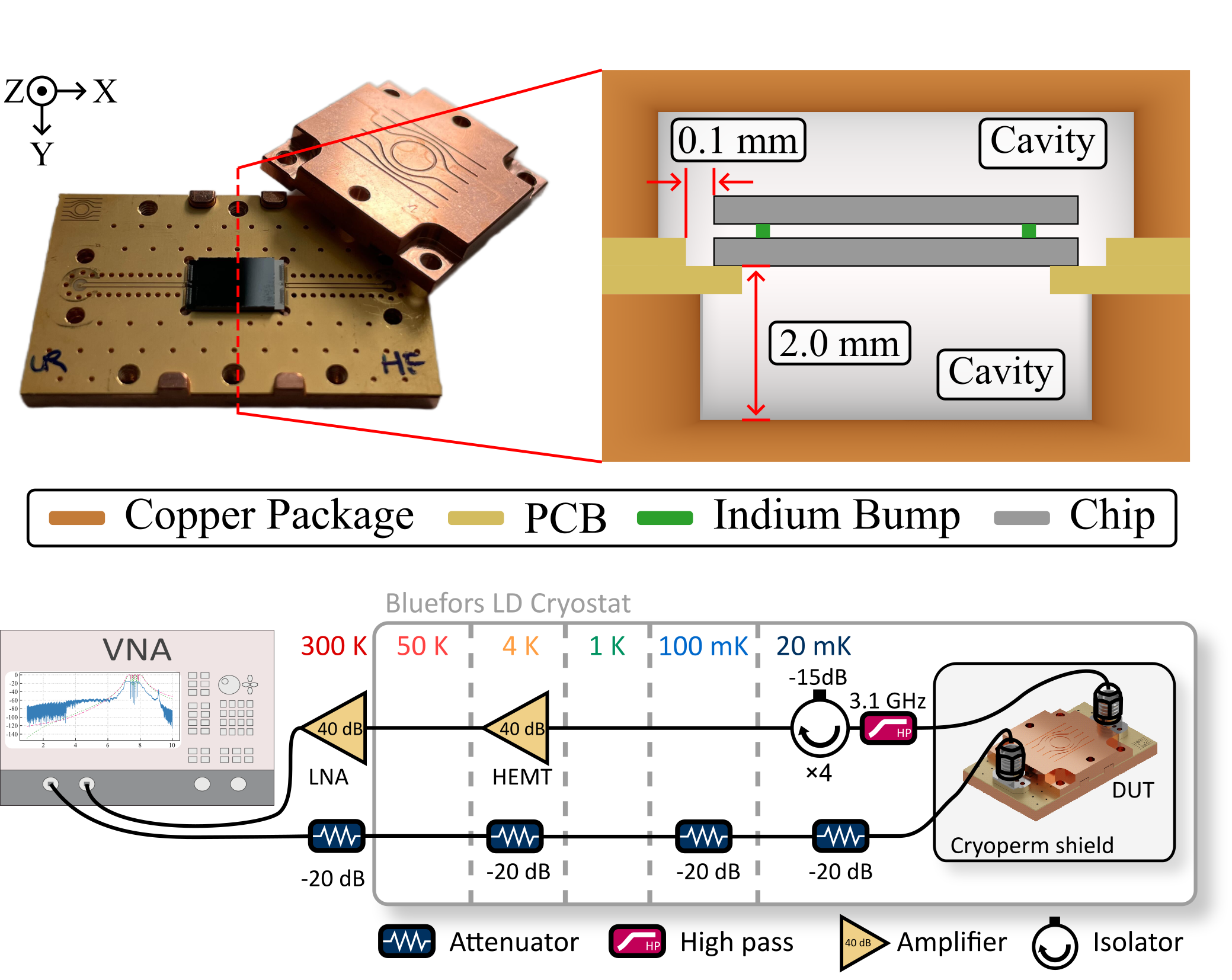}
        \put(0, 76){\normalsize (a)}
        \put(48, 76){\normalsize (b)}
        \put(0, 31){\normalsize (c)}
    \end{overpic}
    \caption{(a) Packaged sample with the lid removed. (b) False-scaled cross-sectional view of the packaged sample in the YZ-plane, illustrating the internal stack-up. (c) Measurement setup for cryogenic characterizations.}
    \label{fig:setup}
\end{figure}
The package consists of a copper base, a carrier printed circuit board (PCB), and a copper lid, as shown in Fig.~\ref{fig:setup}(a). The fabricated device is mounted within an aperture in the carrier PCB and is wire-bonded to the CPW traces on the PCB, which are further connected to the measurement setup via SMA connectors. Fig.~\ref{fig:setup}(b) illustrates the cross-sectional view in the YZ-plane, showing the internal stack-up of the packaged structure. Both the copper lid and the copper base include milled cavities above and below the chip, respectively.

The packaged sample is magnetically shielded with a cryoperm shield and placed in a commercial dilution fridge (Bluefors LD 400 with a base temperature of around 20\,mK). Vector network analyzer (\textit{Keysight N5224B}) is used to perform spectroscopy. The input line contains 80\,dB attenuators in total at different temperature stages. The output line includes a cryogenic HEMT amplifier (\textit{LNF LNC4\_8C}) at 4\,K stage, 2 dual junction isolators (\textit{LNF-ISISC4\_12A}), and a 3.1\,GHz high pass filter (\textit{Minicircuits VHF-3100+}) at 20\,mK stage. The detailed measurement setup is presented in Fig.~\ref{fig:setup}(c).

% =======
% Tabel 2
% =======
{
\begin{table*}[t]
\setlength\tabcolsep{12pt}
\caption{Geometric parameters of readout resonators, along with the corresponding resonance frequencies and effective external $Q$-factors obtained from HFSS driven-modal simulation and the proposed analytical model.}
\renewcommand{\arraystretch}{1.2}
\centering
\normalsize
\begin{tabular}{l  c  c  c  c  c  c}
\hline
\textbf{Parameter} & \textbf{Res. 1} & \textbf{Res. 2} & \textbf{Res. 3} & \textbf{Res. 4} & \textbf{Res. 5} & \textbf{Res. 6} \\
\hline
Total length, $l_t$ ($\mu\text{m}$) & 4400 & 4350 & 4300 & 4250 & 4200 & 4150\\
Short section length, $l_s$ ($\mu\text{m}$) & 500 & 500 & 500 & 500 & 500 & 500\\
Coupling section length, $l_c$ ($\mu\text{m}$) & 300 & 200 & 200 & 200 & 200 & 300\\
Coupling section spacing $d$ ($\mu\text{m}$) & 7 & 10 & 15 & 15 & 10 & 7\\
Coupling section location $l_{c,2}$\textsuperscript{a} ($\mu\text{m}$) & 2367 & 1917 & 1417 & 1417 & 1917 & 2367\\
\hline
Resonance frequency, $f_r^{\text{sim.}}$ (GHz)& 7.42 & 7.54 & 7.65 & 7.74 & 7.82 & 7.90 \\
Resonance frequency, $f_r^{\text{cal.}}$ (GHz)& 7.43 & 7.51 & 7.60 & 7.69 & 7.78 & 7.87 \\
External Q-factor, $Q_e^{\text{sim,}}$ ($\times10^3$)& 1.36 & 1.39 & 1.70 & 1.72 & 1.30 & 1.35\\
External Q-factor, $Q_e^{\text{cal,}}$ ($\times10^3$)& 0.91 & 1.07 & 1.57 & 1.98 & 2.40 & 1.18\\
\hline
\end{tabular}
\par\vspace{2pt}
{\begin{minipage}{0.9\textwidth}
\raggedright 
\footnotesize
\textsuperscript{a}The relative position of the coupling section counting from the common short-end of the CPW line resonators.\par
\end{minipage}}
\label{tab:gep_resonators}
\end{table*}
}

\subsection{\label{sec:Results} Measurement Results and Discussion}
\begin{figure}[!t]
    \centering
    \begin{overpic}{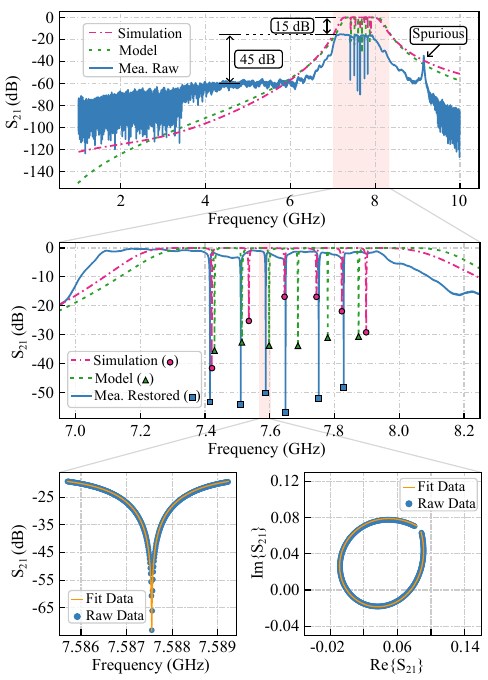}
        \put(0, 97){\normalsize (a)}
        \put(0, 63){\normalsize (b)}
        \put(0, 29){\normalsize (c)}
        \put(35, 29){\normalsize (d)}
    \end{overpic}
    \caption{(a) Measured $S_{21}^{\text{mea.}}$ (blue) spectrum from $1$ to $10\,\text{GHz}$, compared with $S_{21}^{\text{cal.}}$ (green) calculated from the analytical model and $S_{21}^{\text{sim.}}$ (magenta) obtained from Ansys HFSS driven-modal simulations.(b) $S_{21}$ spectrum with smaller window from 7.0 to 8.2\,GHz. The measured $S_{21}^{\text{mea.}}$ is restored by compensating for a $15\,\text{dB}$ loss in the signal path. Markers are added to indicate the resonance frequencies of readout resonators for the analytical model ($\triangle$), simulation ($\circ$), and measurement ($\square$). (c) Measured raw data (blue) and fitted $S_{21}$ response (orange) of a representative readout resonator (Res.~3). (d) Corresponding complex-plane representation of $S_{21}$ for the same resonator, showing $\text{Re}\{S_{21}\}$ and $\text{Im}\{S_{21}\}$.}
    \label{fig:spectrum}
\end{figure}

The raw measurement data of transmission spectrum $S_{21}^{\text{mea.}}$, shown in Fig.~\ref{fig:spectrum}(a), spans 1-10\,GHz with 100~001 points at a VNA output power of 0\,dBm. The calculated ($S_{21}^{\text{cal.}}$) and simulated ($S_{21}^{\text{sim.}}$) responses are included for comparison. As the results shown, the passband exhibits an overall amplitude reduction of approximately -15\,dB compared with the analytical model and simulation. This behavior is consistent with our prior measurements on the same 3D flip-chip platform~\cite{luo2025} and is mainly attributed to ohmic losses along the signal path in the measurement setup. To compensate for this attenuation, a flat +15\,dB magnitude correction is applied to the measured raw date. The restored spectrum, shown in Fig.~\ref{fig:spectrum}(b) over the smaller frequency span (7.0-8.2\,GHz), closely tracks the modeling result $S_{21}^{\text{cal.}}$ and the simulation $S_{21}^{\text{sim.}}$ obtained by Ansys HFSS (driven-modal) across the passband. All six floating readout resonances appear within the passband in the three traces (model, simulation, and measurement) as indicated by the markers in Fig.~\ref{fig:spectrum}(b), confirming that the multiplexed configuration is preserved on chip.

% =======
% FIG. 8
% =======

The fabricated device shows a small downshift of the passband center frequency by approximately 200\,MHz relative to the design target, which we attribute to the kinetic inductance of the thin-film Nb. This contribution is additive to the total inductance and modifies the resonance frequency in~(\ref{eq:bare_f}) to
\begin{equation}\label{eq:bare_f_lk}
f_{r} = \frac{1}{4 l_t \sqrt{(L_l^k + L_l^g) C_l^g}}.
\end{equation}
Following the model in~\cite{Keiji1992}, the kinetic inductance per unit length $L_l^k$ is determined by the superconducting material properties, in particular the magnetic penetration depth~\cite{Shima2025}, as well as the current distribution across the conductor cross-section~\cite{Amini2022}. For the 150\,nm Nb film used in this work, the magnetic penetration depth is approximately 80\,nm according to~\cite{Gubin2005, Pinto_2018}. Using this value together with the current distribution of the specific CPW geometry (Table~\ref{tab:geo_filter}) obtained from HFSS simulation, the kinetic inductance per unit length is estimated to be $L_l^k\approx$ 12\,nH/m. The geometrical inductance $L_l^g$, calculated using conformal mapping techniques as described in Appendix~\ref{app:TGCPW}, is approximately 388\,nH/m. This yields a ratio $L_l^k/L_l^g \approx$ 3.1\,\%, corresponding to an expected frequency reduction of approximately 2\,\%. This correction directly affects the self-coupling coefficients $m_{ii}$ in~(\ref{eq:m_matrix}), as they are determined by the resonance frequencies of the individual resonators. While this quantitative estimation applies to CPW-based $\lambda/4$ resonators operating in a quasi-TEM mode, the spiral resonators do not strictly satisfy this condition. Nevertheless, the kinetic inductance remains additive to the total inductance and therefore leads to a similar qualitative downshift in their resonance frequencies.

The Purcell suppression is estimated to be around 45\,dB by comparing $S_{21}(f_r)$ at a representative readout frequency $f_r$ in the passband to $S_{21}(f_q)$ at a typical qubit frequency $f_q\approx4.5\,\text{GHz}$. As shown in Fig.~\ref{fig:spectrum}(a), the proposed filter exhibits a relatively flat $S_{21}$ transmission response over a broader frequency range around 4.5\,GHz (approximately 4–6\,GHz), indicating that the Purcell suppression is robust against variations in qubit frequency arising from fabrication tolerances in the Josephson junctions. 

The measurement $S_{21}^{\text{mea.}}$ exhibits another two notable deviations from the analytical model and the chip-level simulation in Fig.~\ref{fig:spectrum}(a), namely reduced Purcell suppression at 4.5\,GHz in the lower stopband and a spurious peak near 9.2\,GHz in the upper stopband. These discrepancies are attributed to parasitic effects associated with wire bonding and packaging, which are not captured in the chip-level simulation. As shown in Fig.~\ref{fig:setup}(b), the finite clearance between the chip and the aperture in the carrier PCB forms a slot structure that can support slot-line modes. Signals outside the passband of the filter are reflected back towards the input port, and a fraction of the reflected energy can couple into the slot-line mode at the chip-to-PCB interface and propagate to the output port. This parasitic transmission path leads to an increased $S_{21}$ level in the lower stopband compared with the chip-level simulation.
In addition, the wire bonds connecting the signal pads on the chip and the carrier PCB introduce parasitic inductance (approximately 1\,nH/mm~\cite{Wenner2011}), resulting in impedance mismatch that effectively defines the boundary condition for the slot-line mode. This gives rise to a resonant response at a specific frequency, manifested as the spurious peak observed near 9.2\,GHz. Similar effects have been reported and analyzed in~\cite{Huang2021, Benjamin2019}.
Both discrepancies are reproduced in full-wave simulations that include the package and wire bonds, as discussed in Appendix~\ref{app:pkg}. The chip-level simulation in Fig.~\ref{fig:spectrum}(a) is retained for direct comparison with the analytical model, as these package-induced effects do not affect the in-band performance of the filter.

Subsequently, from the measured $S_{21}^{\text{mea.}}$, we extract for each readout resonator the resonance frequency $f_r$, loaded quality factor $Q_l$, and external quality factor $Q_e$ by fitting the complex notch response to the model~\cite{Probst2015}:
\begin{equation}\label{eq:S21_notch_noise}
S_{21}(f) = ae^{j\alpha}e^{-j2\pi f\tau}\left[1-\frac{(Q_l/Q_e)e^{j\phi}}{1+2jQ_l(f/f_r-1)}\right]
\end{equation}
Here, the model also accounts for environmental effects such as attenuation a, electric delay $\alpha$ and $\tau$ introduced by the environment and RF cables, as well as the impedance mismatch $\phi$ in the feedline. The measured data and corresponding fitting results of Res.~3 as a representative example are shown in Fig.~\ref{fig:spectrum}(c) and (d). The internal quality factor $Q_i$ can also be estimated from the fitting model, where the loaded quality factor $Q_l$ is related to the external and internal quality factors through $1/Q_l = 1/Q_e + 1/Q_i$. For Res.~3, $Q_i$ is approximately $2\times10^6$ at a VNA output power of 0\,dBm, which is consistent with typical Nb-based CPW resonators in 3D flip-chip architectures \cite{Norris2024}. However, in the present design the readout resonators are intentionally operated in a strongly overcoupled regime ($Q_e \ll Q_i$) to enable fast readout, which limits the sensitivity to internal loss, as discussed in~\cite{Goppl2008, Gao2008}.

% =======
% FIG. 8
% =======
\begin{figure}[!b]
    \centering
    \begin{overpic}{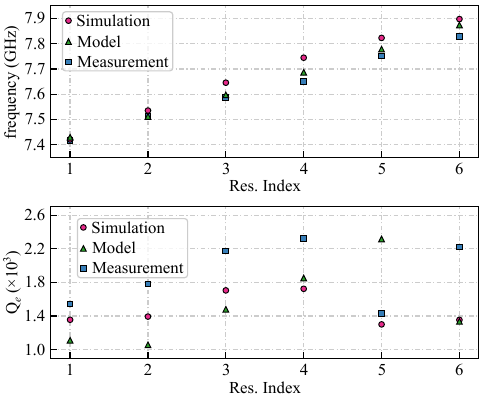}
        \put(-4, 80){\normalsize (a)}
        \put(-4, 38){\normalsize (b)}
    \end{overpic}
    \caption{Resonance frequencies $f_r$ (a) and external Q-factors $Q_e$ (b) extracted from the transmission coefficients obtained by experimental measurements, Ansys HFSS driven-modal simulations and the analytical Model.}
    \label{fig:fr_Qe}
\end{figure}

When applying the notch-type model to $S_{21}$ obtained from simulation or analytical model, those parameters from environmental effects can be ignored and the model in~(\ref{eq:S21_notch_noise}) will be simplified as~\cite{Probst2015}:
\begin{equation}\label{eq:S21_notch}
S_{21}(f) = 1-\frac{Q_l/Q_e}{1+2jQ_l(f/f_r-1)}.
\end{equation}
Fig.~\ref{fig:fr_Qe} compares parameters extracted from the analytical model, HFSS driven-modal simulation, and measurement. As shown in Fig.~\ref{fig:fr_Qe}(a), the measured resonance frequencies differ from the simulated values by $\Delta f=|f_{r}^{\text{mea.}} - f_{r}^{\text{sim.}}|<100\,\text{MHz}$ across all six readout resonators. This systematic downshift is attributed to the kinetic inductance of Nb thin-film, which is not included in the simulation. The same mechanism applies here, as described above in (\ref{eq:bare_f_lk}).

In contrast, the extracted external quality factors exhibit more noticeable discrepancies ($\Delta Q_e = |Q_e^{\text{mea.}}-Q_e^{\text{sim.}}|<0.5\times10^3$, excepting resonator 6) compared to the simulation, corresponding to a relative deviation of approximately $\Delta Q_e/Q_e^{\text{sim.}}\approx30\,\%$. We attribute this primarily to fabrication tolerances in the flip-chip bonding process. Even with the use of polymer spacers, the chip spacing $h_s$ varies across the chip with a mean value of $9.91 \pm 0.12\,\mu\text{m}$. A detailed characterization of $h_s$, performed using a profilometer and presented in Appendix~\ref{app:hs}, shows a systematic variation across the chip. As a result, the local spacing in the coupling sections is smaller than the target design value 10\,$\mu\text{m}$, which introduces nontrivial deviations in the extracted $Q_e$ \cite{luo2025}. In addition, the measured external Q-factors $Q_e^{\text{mea.}}$ of all readout resonators are consistently higher than the simulated values, providing independent evidence of reduced local spacing in the coupling regions, as a closer grounding plane weakens the coupling strength. Exceptionally, resonator~6 exhibits a larger deviation, $\Delta Q_e(\text{Res.6}) \approx 0.9\times10^3$, further indicating non-uniformity of $h_s$ across the sample.

We also observe an offset between the extracted $Q_e$ from the analytical model and simulation. Several factors contribute to this deviation. First, the open ends of the CPW line resonator~2 and 3 in the filter include capacitive patches. Although their effects are incorporated as a length extension in~(\ref{eq:l_eff}), this approximation does not accurately capture the effective position of the coupling section along the $\lambda/4$ line and consequently affect the coupling coefficient as shown in Fig.~\ref{fig:Model_vs_sim}. Second, it is difficult to effectively include the curved bend in the joints between coupling section and the rest parts of resonator shown in Fig.~\ref{fig:main}(f) into the analytical model. Third, weak cross-couplings between the neighboring resonators are neglected in the coupling-matrix formulation in~(\ref{eq:m_matrix}). Despite these limits, the model predicts $Q_e$ within a reasonable margin for all six resonators ($|Q_e^{\text{cal.}}-Q_e^{\text{sim.}}|<0.45\times10^3$ relative to the simulated values), which is sufficient for design iteration and targeting.

\section{Conclusion and Outlook}
In this work, we have demonstrated a compact broadband Purcell filter composed of four coupled resonators integrated on a 3D flip-chip platform. The filter exhibits a flat 1\,GHz passband centered at 7.68\,GHz and strong stopband attenuation exceeding 45\,dB at typical qubit frequencies, ensuring effective Purcell protection. Six floating readout resonators are strongly coupled within the passband, demonstrating fast and efficient readout. An analytical framework has been developed not only to enable rapid circuit synthesis of the Purcell filter but also to fast determine the resonance frequencies and external Q-factors of the readout resonators. 

The proposed design offers several important advantages for large-scale superconducting quantum processors. The over-the-air capacitive coupling between the spiral and CPW line resonators shows high tolerance to fabrication variations, particularly to chip-spacing nonuniformities introduced during bonding process. Its compact footprint and flip-chip compatibility mitigate the routing congestion in large-scale quantum chip and make the design more suitable for dense, multiplexed qubit readout architectures.

While the present work does not include a fully integrated qubit demonstration, the proposed filter design provides a scalable and robust solution for suppressing Purcell decay in multiplexed readout architectures. Future work will focus on integrating the filter with qubit devices to further validate its performance in realistic quantum processor environments.

\section*{Acknowledgments}
The authors gratefully acknowledge fruitful discussion with Kevin Kiener and Saya Schöbe from the Quantum Computing group at the Walther Meissner Institute. The authors gratefully acknowledge Johannes Schirk from the Quantum Computing group at the Walther Meissner Institute for his contributions to the package design.

{\appendices
\section{\label{app:TGCPW} Conformal Mapping for CPW Line with Top Ground}
A top-grounded CPW (TGCPW) line can be analyzed under the quasi-TEM approximation and decomposed to two partial regions. The upper part is in an air cavity seeing the continuous top ground with finite distance $h_s$ and the bottom part is in a Silicon dielectric with finite height $h_b$. Detailed conformal mapping steps are given in \cite{Li2023}. The total impedance can be treated as the parallel combination of those two regions and the closed-form expressions are given in \cite{Gevorgian1995} as:
\begin{gather}
\varepsilon_{\mathrm{eff}, r} = 1+(\varepsilon_r-1)\frac{K(k_{r,3})}{K'(k_{r,3})}\frac{1}{\frac{K(k_{r,1})}{K'(k_{r,1})}+\frac{K(k_{r,2})}{K'(k_{r,2})}}, \label{eq:eps_cpw}\\
Z_r = \frac{1}{c\varepsilon_0\sqrt{\varepsilon_{\mathrm{eff}, r}}\left(\frac{K(k_{r,1})}{K'(k_{r,1})}+\frac{K(k_{r,2})}{K'(k_{r,2})}\right)}, \label{eq:Z0_cpw}
\end{gather}
where
\begin{align*}
    k_{r,1} &= w/(w+2g), \\
    k_{r,2} &= \tanh(\pi w/4h_s)/\tanh(\pi (w+2g)/4h_s), \\
    k_{r,3} &= \sinh(\pi w/4h_b)/\sinh(\pi (w+2g)/4h_b).
\end{align*}

\section{\label{app:coupler} Analytical Model for Coupled-line Coupler}
The coupled-line section is modeled as a symmetrical four-port network described by its impedance matrix, which relates the port voltages and currents as:
\begin{equation}\label{eq:coupler_z_matrix}
\begin{bmatrix}
    V_{1} \\
    \vdots\\
    V_{4}
\end{bmatrix}
=
\begin{bmatrix}
    Z_{11} & ... & Z_{14} \\
    \vdots & \ddots &\vdots\\
    Z_{41} & ... & Z_{44}    
\end{bmatrix}
\begin{bmatrix}
    I_{1} \\
    \vdots\\
    I_{4} \\
\end{bmatrix}.
\end{equation}
The entries of the impedance matrix in~(\ref{eq:coupler_z_matrix}) follow directly from even- and odd-mode analysis of the coupled transmission lines as:
\begin{equation}\label{eq:z_entries}
\begin{aligned}
    Z_{11}=Z_{22}=Z_{33}=Z_{44} =  \frac{-j}{2}(Z_{0,e} + Z_{0,o}) \cot\theta \\
    Z_{12}=Z_{21}=Z_{34}=Z_{43} =  \frac{-j}{2}(Z_{0,e} - Z_{0,o}) \cot\theta \\
    Z_{13}=Z_{31}=Z_{24}=Z_{42} =  \frac{-j}{2}(Z_{0,e} - Z_{0,o}) \csc\theta \\
    Z_{14}=Z_{41}=Z_{23}=Z_{32} =  \frac{-j}{2}(Z_{0,e} + Z_{0,o}) \csc\theta \\
\end{aligned}
\end{equation}
Here $Z_{0,e}$ and $Z_{0,o}$ denote the characteristic impedances of even and odd modes, respectively. The electrical length $\theta$ depends on the effective dielectric constant and the physical length of the coupling section $l_c$.

\section{\label{app:Coupled_TGCPW} Conformal Mapping for Edge-coupled TGCPW Line}
Fig.~\ref{fig:CCPW_Efield} shows the electric-field distribution in the cross section of an edge-coupled TGCPW pair simulated in Ansys 2D Extractor. A virtual perfect magnetic conductor (PMC) or perfect electric conductor (PEC) plane can be introduced along the symmetry plane (BB' plane in Fig.~\ref{fig:main}(a)) between the two conductor strips without perturbing the field distribution. Due to the symmetry, only half of the geometry needs to be analyzed for either mode.

The geometries in Fig.~\ref{fig:CCPW_Efield} are defined by the coordinates:
\begin{equation}\label{eq:point_in_z}
    \begin{aligned}
        z_a &= d/2, \\
        z_b &= d/2 + w, \\
        z_c &= d/2 + w + g,\\
    \end{aligned}
\end{equation}

The total capacitance per-unit-length of the edge-coupled TGCPW pair can be decomposed into three partial capacitances that represent the distinct dielectric regions involved as
\begin{equation}\label{eq:C_tot}
    C_{tot,m} = C_{a,m1} + C_{a,m2} + C_{d,m}
\end{equation}
where \textit{m} represents the propagation mode, either even (\textit{e}) or odd (\textit{o}). As illustrated in Fig.~\ref{fig:CCPW_Efield}, $C_{a,m1}$ corresponds to the capacitance in the upper air region within the flip-chip cavity, $C_{a,m2}$ represents the capacitance in the lower region in the absence of the dielectric substrate, and $C_{d,m}$ denotes the capacitance associated with the silicon substrate.

% =======
% FIG. 9
% =======
\begin{figure}[t]
    \centering
    \begin{overpic}{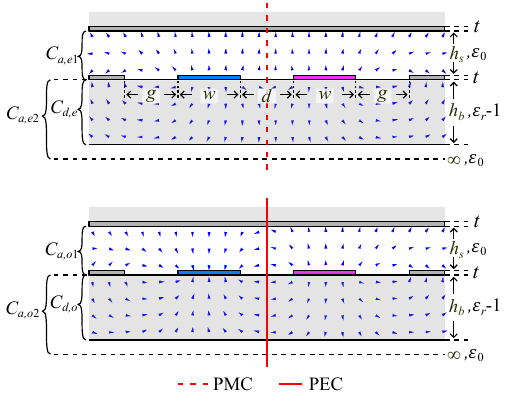}
        \put(0, 75){\normalsize (a)}
        \put(0, 38){\normalsize (b)}
    \end{overpic}
    \caption{Cross-sectional E-field distributions for edge-coupled TGCPW line simulated using Ansys 2D Extractor, and the corresponding partial capacitance configurations for even mode (a) and odd mode (b).}
    \label{fig:CCPW_Efield}
\end{figure}

In region~1, the top ground provides anther PEC boundary condition with a finite distance, $h_s$, to the structure. Therefore, the partial capacitances $C_{a, m1}$ in this air cavity are \cite{Hanna1985}:
\begin{equation}\label{eq:C_a1}
    \begin{gathered}
    C_{a,e1} = 2\varepsilon_0\frac{K(k_e)}{K'(k_e)},\\
    C_{a,o1} = 2\varepsilon_0\frac{K(k_o)}{K'(k_o)},    
    \end{gathered}
\end{equation}
where 
\begin{align*}
    k_e &= \phi_1\frac{-(\phi_1^2-\phi_2^2)^{1/2}+(\phi_1^2-\phi_3^2)^{1/2}}{\phi_3(\phi_1^2-\phi_2^2)^{1/2}+\phi_2(\phi_1^2-\phi_3^2)^{1/2}}\\
    k_o &= \phi_4\frac{-(\phi_4^2-\phi_5^2)^{1/2}+(\phi_4^2-\phi_6^2)^{1/2}}{\phi_6(\phi_4^2-\phi_5^2)^{1/2}+\phi_5(\phi_4^2-\phi_6^2)^{1/2}}\\
    \phi_1 &=\frac{1}{2}\text{cosh}^2\left(\frac{\pi z_c}{2h_s}\right),\\
    \phi_2 &= \text{sinh}^2\left(\frac{\pi z_b}{2h_s}\right)-\phi_1+1,\\
    \phi_3 &=  \text{sinh}^2\left(\frac{\pi z_a}{2h_s}\right)-\phi_1+1,\\
\end{align*}
\begin{align*}
    \phi_4 &=\frac{1}{2}\text{sinh}^2\left(\frac{\pi z_c}{2h_s}\right),\\
    \phi_5 &= \text{sinh}^2\left(\frac{\pi z_b}{2h_s}\right)-\phi_4,\\
    \phi_6 &=  \text{sinh}^2\left(\frac{\pi z_a}{2h_s}\right)-\phi_4.
\end{align*}

In region~2, the fields extend into an effectively infinite air region in the absence of dielectric. The corresponding partial capacitances $C_{a, m2}$ are therefore given by \cite{Hanna1985}
\begin{equation}\label{eq:C_a2}
    \begin{gathered}
    C_{a,e2} = \varepsilon_0\frac{K(k_{e2})}{K'(k_{e2})},\\
    C_{a,o2} = \varepsilon_0\frac{K(k_{o2})}{K'(k_{o2}))},    
    \end{gathered}
\end{equation}
where 
\begin{align*}
    k_{e2} = \sqrt{\frac{z_b^2-z_a^2}{z_c^2-z_a^2}}, \quad k_{o2} = \sqrt{\frac{z_c^2(z_b^2-z_a^2)}{z_b^2(z_c^2-z_a^2)}}.\\
\end{align*}

In region~3, the structure is backed by a dielectric substrate with finite distance, $h_b$, to the dielectric-to-air boundary. The coordinates in~(\ref{eq:point_in_z}) after conformal mapping are transformed according to
\begin{equation}\label{zs_points}
    z_{n,h} = \sinh(\pi z_n/2h_b), \quad n=a, b, c.
\end{equation}
The corresponding partial capacitance $C_{d,m}$ can then be written as \cite{Cheng1995}
\begin{equation}\label{eq:C_d}
    \begin{gathered}
    C_{d,e} = \varepsilon_0 (\varepsilon_r-1)\frac{K(k_{e3})}{K'(k_{e3})},\\
    C_{d,o} = \varepsilon_0 (\varepsilon_r-1)C_p(\frac{W}{H}, \frac{W_1}{W}, \frac{W_2}{W}), \\
    \end{gathered}
\end{equation}
where 
\begin{align*}
    k_{e3} &= \sqrt{\frac{z_{b,h}^2-z_{a,h}^2}{z_{c,h}^2-z_{a,h}^2}},\\
    k_{o3} &= \sqrt{\frac{z_{c,h}^2(z_{b,h}^2-z_{a,h}^2)}{z_{b,h}^2(z_{c,h}^2-z_{a,h}^2)}},\\
    W &= K(k_{o3}), H = K'(k_{o3}),\\
    W_1 &= F\left(\arcsin\sqrt{\frac{z_{c,h}^2-z_{a,h}^2}{z_{c,h}^2\text{cosh}^2(\pi z_a/2h_b)}}, k_{o3}\right), \\
    W_2 &= F\left(\arcsin\sqrt{\frac{z_{c,h}^2-z_{a,h}^2}{z_{c,h}^2}}, k_{o3}\right). \\
\end{align*}

Here, $C_p$ is a function of three geometric parameters, $\alpha,\beta,\gamma$, which define the configuration of a parallel-plate capacitor with a infinite long slot on one plate. The analytical evaluation of $C_p$ follows the conformal mapping approach described in \cite{Cheng1995, Cheng1997}. Some useful closed-form relations for solving~(\ref{eq:C_d}) are given as
\begin{align*}
    &C_p(\alpha,\beta,\gamma) = K(k_1)/K(k_1') + K(k_3)/K(k_3'), \\
    &\delta = \frac{(\beta+\gamma)}{2}, \\
    &\frac{K(k_4)}{K(k_4')} = \alpha(1 - \delta), \\
    &\frac{K(k_2)}{K(k_2')} = \alpha\delta, \\
    &\frac{F[\arcsin(k_3/k_4), k_4]}{K(k_4)} = \frac{1-\gamma}{1-\delta}, \\
    &\frac{F[\arcsin(k_1/k_2), k_2]}{K(k_2)} = \frac{\beta}{\delta}.
\end{align*}

By performing the conformal mapping procedure for each region, all the terms in (\ref{eq:C_tot}) have been determined using closed-form expressions. Consequently, the effective dielectric constants for both the even and odd modes, $\varepsilon_{\mathrm{eff}, m}$, and the characteristic impedances $Z_{0,m}$ are calculated using:
\begin{gather}
\varepsilon_{\mathrm{eff}, m} = \frac{C_{tot,m}}{C_{a,m}} = \frac{C_{tot,m}}{C_{a,m1}+C_{a,m2}}, \label{eq:eps_eff}\\
Z_{0,m} = \frac{1}{c_0 C_{a,m} \sqrt{\varepsilon_{\mathrm{eff}, m}}}, \label{eq:Z_0m}
\end{gather}
where $c_0$ is the speed of light in free space. By solving $Z_{0,e}$ and $Z_{0,o}$ and substituting the results into (\ref{eq:z_entries}), the impedance matrix of the coupled TGCPW line in the flip-chip configuration, $\mathbf{Z}^{\text{coupler}}$, is obtained. The corresponding scattering matrix, $\mathbf{S}^{\text{coupler}}$, is then derived via standard impedance-to-scattering transformation and is used in Section~\ref{sec:M_ij} to model the coupling behavior between adjacent resonators.

To validate the accuracy of the conformal mapping approach, we apply it to an edge-coupled symmetric TGCPW structure and compare the calculated results from~(\ref{eq:Z_0m}) with those obtained from Ansys Q2D Extractor simulations. Two metallization conditions are considered: an ideal case with zero conductor thickness, $t=0\,\text{nm}$ and a practical case with metal thickness of $t=150\,\text{nm}$, corresponding to the thin-film Nb used in fabrication. For both cases, the TGCPW geometry is defined by conductor width $w=10\,\mu\text{m}$ and gap $g=9\,\mu\text{m}$. Chip spacing is fixed as $h_s=10\,\mu\text{m}$. The edge-to-edge spacing \textit{d} between the coupled strips is swept from 1 to 15\,$\mu\text{m}$ to evaluate the dependence of the coupling parameters on line separation.

% =======
% FIG. 10
% =======
\begin{figure}[h]
    \includegraphics{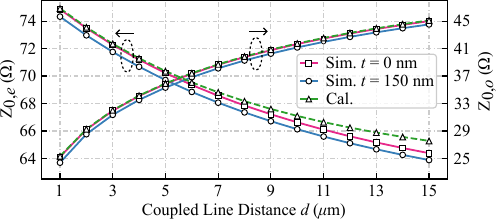}%
    \caption{Calculated characteristic impedance for even and odd mode, comparing with Ansys Q2D simulations under different configurations.}
    \label{fig:conformal_mapping}
\end{figure}

As shown in Fig.~\ref{fig:conformal_mapping}, the characteristic impedances $Z_{0,m}$ of even and odd modes calculated from the conformal mapping model agree closely with the results from the Q2D simulations. Defining the relative error as
\begin{equation}
    \mathrm{err} = \frac{|Z_{0, m}^\mathrm{Cal.} - Z_{0, m}^\mathrm{Sim.}|}{Z_{0, m}^\mathrm{Sim.}},
\end{equation}
the maximum deviation between the calculated and simulated values for the case with $t=0\,\text{nm}$ remains within 1.3\,\%. The error increases slightly when comparing to the case with $t=150\,\text{nm}$. The discrepancy arises primarily from the finite metal thickness, which is not included in the conformal mapping formulation. This inherent approximation of the conformal mapping method propagates into subsequent calculations, such as the estimation of coupling coefficients.

\section{\label{app:pkg} Packaging Effects}
\begin{comment}
    \textcolor{red}{In Fig.~\ref{fig:spectrum}(a), two main discrepancies are observed between the simulated $S_{21}^{\text{sim.}}$ and measured $S_{21}^{\text{mea.}}$ transmission responses. At the qubit frequency $f_q\approx$ 4.5\,GHz, the measured Purcell suppression is approximately 45\,dB, which is lower than the simulated value. Second, a spurious resonance is observed near 9.2\,GHz in the upper stopband. Since $S_{21}^{\text{sim.}}$ is obtained from chip-level simulations, these discrepancies are attributed to packaging effects.}
\end{comment}

To investigate the impact of packaging effects discussed in Section~\ref{sec:Results}, full-wave electromagnetic simulations including the package are performed. The package configuration follows that described in Section~\ref{sec:Setup}. Aluminum wire bonds with a length of 0.75\,mm and a diameter of 25\,$\mu\text{m}$ are included for both signal and ground connections. The floating readout resonators are removed from the simulation model to simplify the analysis.

Fig.~\ref{fig:s21_pkg} presents the simulated transmission response $S_{21}^{\text{sim.}}$ with and without package. Including the package leads to two key changes. First, the transmission level in the lower stopband increases, reducing Purcell suppression to approximately 62\,dB at 4.5\,GHz, in closer agreement with measurements. Second, an additional spurious resonance appears around 9.1\,GHz, consistent with the measured response in the upper stopband. These results confirm that the discrepancies observed in the measurement can be attributed to packaging-induced parasitic effects.

% =======
% FIG. 12
% =======
\begin{figure}[!t]
    \includegraphics{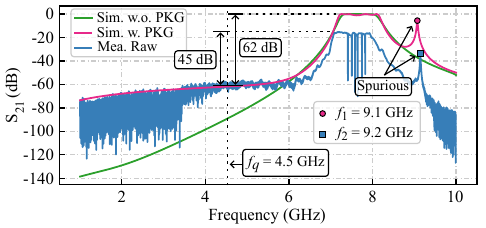}%
    \caption{Simulated transmission response $S_{21}^{\text{sim.}}$ of the filter without readout resonators, with and without the package. Measurement result $S_{21}^{\text{mea.}}$ is included for comparison.}
    \label{fig:s21_pkg}
\end{figure}

% =======
% FIG. 13
% =======
\begin{figure}[!b]
    \centering
    \begin{overpic}{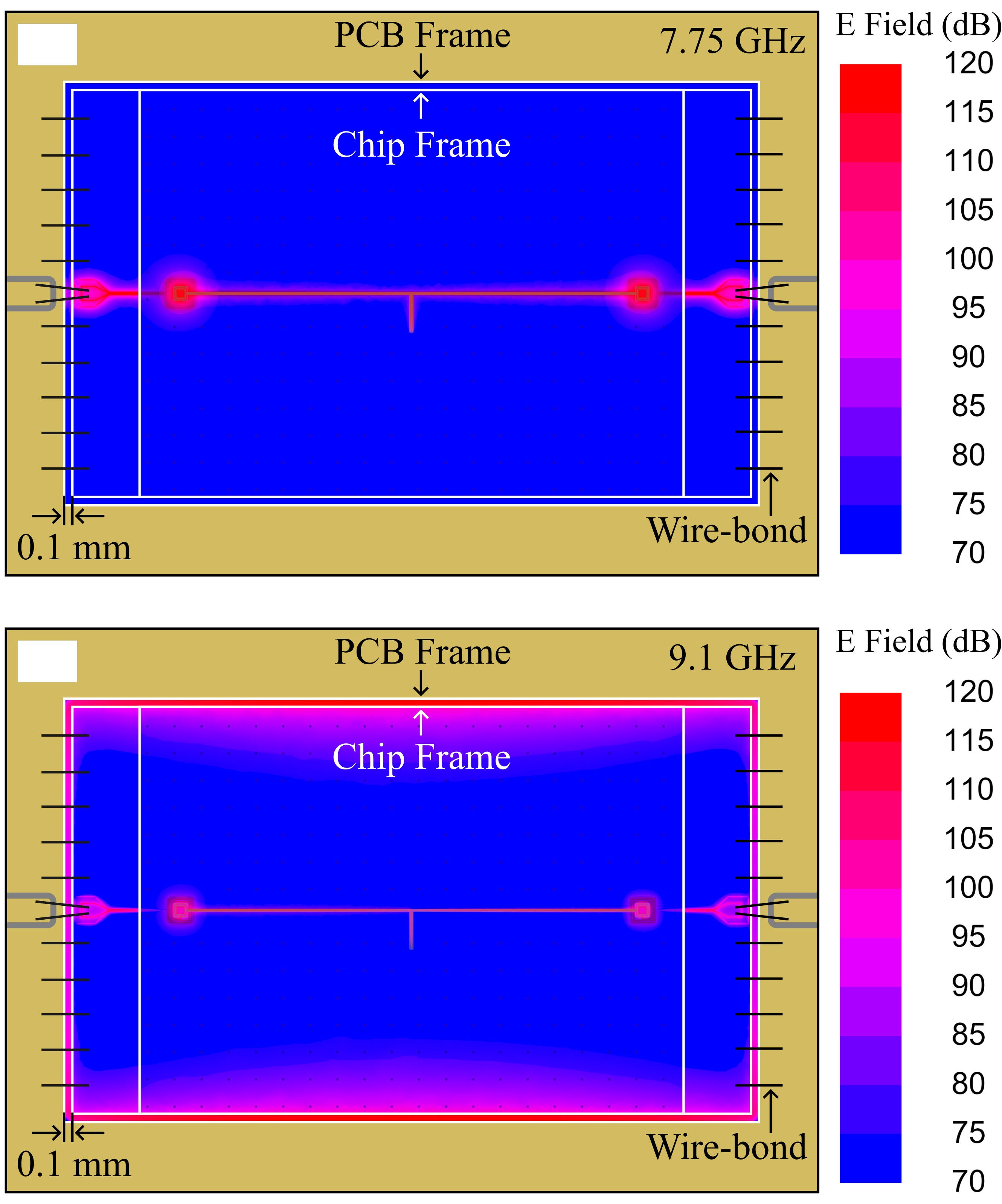}
    \put(1.8, 95.3){\normalsize (a)}
    \put(1.8, 44.2){\normalsize (b)}
    \end{overpic}
    \caption{Electric field distribution in the XY-plane at (a) 7.75\,GHz and (b) 9.1\,GHz. The color scales are identical in both plots for direct comparison.}
    \label{fig:Efield_pkg}
\end{figure}

To further support this interpretation, the electric field distributions at representative frequencies are examined. Fig.~\ref{fig:Efield_pkg}(a) shows the field distribution at 7.75\,GHz within the passband, where the electromagnetic field is primarily confined along the intended signal path. In contrast, Fig.~\ref{fig:Efield_pkg}(b) shows the field distribution at 9.1\,GHz, where strong field localization is observed along the chip-to-PCB interface, indicating the excitation of a parasitic mode.

It should be noted that the simulations are based on ideal conditions. In practice, variations in wire bond geometry and slight misalignment of the chip may lead to deviations from the simulated response. Therefore, the simulations presented here are intended to identify the origin of the spurious resonance rather than to quantitatively reproduce the measured response.

\section{\label{app:hs} Flip-chip Bonding Characterization}
The flip-chip spacing across the chip is characterized using profilometer, and the measured data are fitted to obtain a three-dimensional contour-map, as shown in Fig.~\ref{fig:hs}. From this analysis, the mean spacing is $9.91 \pm 0.12\,\mu\text{m}$, with a minimum of 9.69\,$\mu\text{m}$ and a maximum of 10.32\,$\mu\text{m}$. The results reveal a systematic spatial variation in the chip spacing across the chip, with reduced spacing near the center. These measurements support the interpretation in Section~\ref{sec:Results} that local spacing in the coupling sections is smaller than the design value, contributing to the observed deviations in the extracted external quality factors.

% =======
% FIG. 14
% =======
\begin{figure}[h]
    \includegraphics{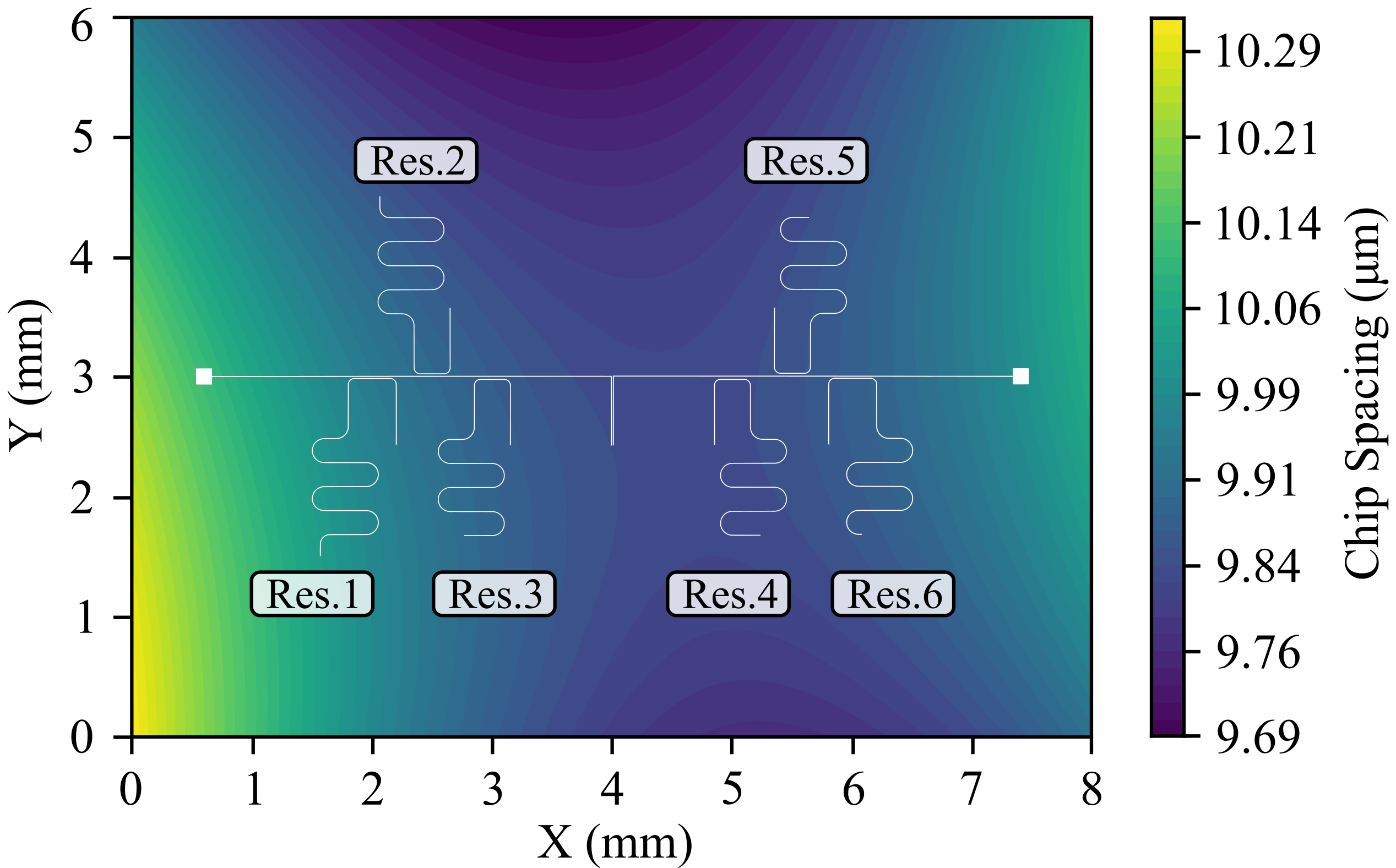}%
    \caption{Contour map of the flip-chip spacing $h_s$ across the chip, obtained by profilometer.}
    \label{fig:hs}
\end{figure}

}

\bibliographystyle{IEEEtran}
\bibliography{IEEEabrv,Bibliography}

@PREAMBLE{
 "\providecommand{\noopsort}[1]{}" 
 # "\providecommand{\singleletter}[1]{#1}%" 
}

@article{Park2024,
    author = {Park, Seong Hyeon and Choi, Gahyun and Kim, Gyunghun and Jo, Jaehyeong and Lee, Bumsung and Kim, Geonyoung and Park, Kibog and Lee, Yong-Ho and Hahn, Seungyong},
    title = {Characterization of broadband Purcell filters with compact footprint for fast multiplexed superconducting qubit readout},
    journal = {Appl. Phys. Lett.},
    volume = {124},
    number = {4},
    pages = {044003},
    year = {2024},
    month = {Jan.},
    issn = {0003-6951},
    doi = {10.1063/5.0182642},
    url = {https://doi.org/10.1063/5.0182642},
}

@article{Jeffrey2014,
  title = {Fast Accurate State Measurement with Superconducting Qubits},
  author = {Jeffrey, Evan and Sank, Daniel and Mutus, J. Y. and White, T. C. and Kelly, J. and Barends, R. and Chen, Y. and Chen, Z. and Chiaro, B. and Dunsworth, A. and Megrant, A. and O'Malley, P. J. J. and Neill, C. and Roushan, P. and Vainsencher, A. and Wenner, J. and Cleland, A. N. and Martinis, John M.},
  journal = {Phys. Rev. Lett.},
  volume = {112},
  issue = {19},
  pages = {190504},
  numpages = {5},
  year = {2014},
  month = {May},
  publisher = {American Physical Society},
  doi = {10.1103/PhysRevLett.112.190504},
  url = {https://link.aps.org/doi/10.1103/PhysRevLett.112.190504}
}

@article{Yan2023,
    author = {Yan, Haoxiong and Wu, Xuntao and Lingenfelter, Andrew and Joshi, Yash J. and Andersson, Gustav and Conner, Christopher R. and Chou, Ming-Han and Grebel, Joel and Miller, Jacob M. and Povey, Rhys G. and Qiao, Hong and Clerk, Aashish A. and Cleland, Andrew N.},
    title = {Broadband bandpass Purcell filter for circuit quantum electrodynamics},
    journal = {Appl. Phys. Lett.},
    volume = {123},
    number = {13},
    pages = {134001},
    year = {2023},
    month = {Sep.},
    issn = {0003-6951},
    doi = {10.1063/5.0161893},
    url = {https://doi.org/10.1063/5.0161893},
    eprint = {https://pubs.aip.org/aip/apl/article-pdf/doi/10.1063/5.0161893/18137590/134001_1_5.0161893.pdf},
}

@article{Reed2010,
    author = {Reed, M. D. and Johnson, B. R. and Houck, A. A. and DiCarlo, L. and Chow, J. M. and Schuster, D. I. and Frunzio, L. and Schoelkopf, R. J.},
    title = {Fast reset and suppressing spontaneous emission of a superconducting qubit},
    journal = {Appl. Phys. Lett.},
    volume = {96},
    number = {20},
    pages = {203110},
    year = {2010},
    month = {May},
    issn = {0003-6951},
    doi = {10.1063/1.3435463},
    url = {https://doi.org/10.1063/1.3435463},
    eprint = {https://pubs.aip.org/aip/apl/article-pdf/doi/10.1063/1.3435463/13988301/203110_1_online.pdf},
}

@article{Krantz2019,
    author = {Krantz, P. and Kjaergaard, M. and Yan, F. and Orlando, T. P. and Gustavsson, S. and Oliver, W. D.},
    title = {A quantum engineer's guide to superconducting qubits},
    journal = {Appl. Phys. Rev.},
    volume = {6},
    number = {2},
    pages = {021318},
    year = {2019},
    month = {Jun.},
    issn = {1931-9401},
    doi = {10.1063/1.5089550},
    url = {https://doi.org/10.1063/1.5089550},
    eprint = {https://pubs.aip.org/aip/apr/article-pdf/doi/10.1063/1.5089550/20722375/021318_1_1.5089550.pdf},
}

@article{Bronn2015,
    author = {Bronn, Nicholas T. and Liu, Yanbing and Hertzberg, Jared B. and Córcoles, Antonio D. and Houck, Andrew A. and Gambetta, Jay M. and Chow, Jerry M.},
    title = {Broadband filters for abatement of spontaneous emission in circuit quantum electrodynamics},
    journal = {Appl. Phys. Lett.},
    volume = {107},
    number = {17},
    pages = {172601},
    year = {2015},
    month = {Oct.},
    issn = {0003-6951},
    doi = {10.1063/1.4934867},
    url = {https://doi.org/10.1063/1.4934867},
    eprint = {https://pubs.aip.org/aip/apl/article-pdf/doi/10.1063/1.4934867/13149681/172601_1_online.pdf},
}

@article{Sunada2022,
  title = {Fast Readout and Reset of a Superconducting Qubit Coupled to a Resonator with an Intrinsic Purcell Filter},
  author = {Sunada, Y. and Kono, S. and Ilves, J. and Tamate, S. and Sugiyama, T. and Tabuchi, Y. and Nakamura, Y.},
  journal = {Phys. Rev. Appl.},
  volume = {17},
  issue = {4},
  pages = {044016},
  numpages = {12},
  year = {2022},
  month = {Apr.},
  publisher = {American Physical Society},
  doi = {10.1103/PhysRevApplied.17.044016},
  url = {https://link.aps.org/doi/10.1103/PhysRevApplied.17.044016}
}

@article{Heinsoo2018,
  title = {Rapid High-fidelity Multiplexed Readout of Superconducting Qubits},
  author = {Heinsoo, Johannes and Andersen, Christian Kraglund and Remm, Ants and Krinner, Sebastian and Walter, Theodore and Salath\'e, Yves and Gasparinetti, Simone and Besse, Jean-Claude and Poto\ifmmode \check{c}\else \v{c}\fi{}nik, Anton and Wallraff, Andreas and Eichler, Christopher},
  journal = {Phys. Rev. Appl.},
  volume = {10},
  issue = {3},
  pages = {034040},
  numpages = {14},
  year = {2018},
  month = {Sep.},
  publisher = {American Physical Society},
  doi = {10.1103/PhysRevApplied.10.034040},
  url = {https://link.aps.org/doi/10.1103/PhysRevApplied.10.034040}
}

@article{Sete2015,
  title = {Quantum theory of a bandpass Purcell filter for qubit readout},
  author = {Sete, Eyob A. and Martinis, John M. and Korotkov, Alexander N.},
  journal = {Phys. Rev. A},
  volume = {92},
  issue = {1},
  pages = {012325},
  numpages = {13},
  year = {2015},
  month = {Jul.},
  publisher = {American Physical Society},
  doi = {10.1103/PhysRevA.92.012325},
  url = {https://link.aps.org/doi/10.1103/PhysRevA.92.012325}
}

@article{Purcell1946,
  title = {Resonance Absorption by Nuclear Magnetic Moments in a Solid},
  author = {Purcell, E. M. and Torrey, H. C. and Pound, R. V.},
  journal = {Phys. Rev.},
  volume = {69},
  issue = {1-2},
  pages = {37--38},
  numpages = {0},
  year = {1946},
  month = {Jan.},
  publisher = {American Physical Society},
  doi = {10.1103/PhysRev.69.37},
  url = {https://link.aps.org/doi/10.1103/PhysRev.69.37}
}

@article{Houch2008,
  title = {Controlling the Spontaneous Emission of a Superconducting Transmon Qubit},
  author = {Houck, A. A. and Schreier, J. A. and Johnson, B. R. and Chow, J. M. and Koch, Jens and Gambetta, J. M. and Schuster, D. I. and Frunzio, L. and Devoret, M. H. and Girvin, S. M. and Schoelkopf, R. J.},
  journal = {Phys. Rev. Lett.},
  volume = {101},
  issue = {8},
  pages = {080502},
  numpages = {4},
  year = {2008},
  month = {Aug.},
  publisher = {American Physical Society},
  doi = {10.1103/PhysRevLett.101.080502},
  url = {https://link.aps.org/doi/10.1103/PhysRevLett.101.080502}
}

@article{Blais2021,
  title = {Circuit quantum electrodynamics},
  author = {Blais, Alexandre and Grimsmo, Arne L. and Girvin, S. M. and Wallraff, Andreas},
  journal = {Rev. Mod. Phys.},
  volume = {93},
  issue = {2},
  pages = {025005},
  numpages = {72},
  year = {2021},
  month = {May},
  publisher = {American Physical Society},
  doi = {10.1103/RevModPhys.93.025005},
  url = {https://link.aps.org/doi/10.1103/RevModPhys.93.025005}
}

@Article{Saxberg2022,
author={Saxberg, Brendan and Vrajitoarea, Andrei and Roberts, Gabrielle and Panetta, Margaret G. and Simon, Jonathan and Schuster, David I.},
title={Disorder-assisted assembly of strongly correlated fluids of light},
journal={Nature},
year={2022},
month={Dec.},
day={01},
volume={612},
number={7940},
pages={435-441},
issn={1476-4687},
doi={10.1038/s41586-022-05357-x},
url={https://doi.org/10.1038/s41586-022-05357-x}
}

@article{Chen2012,
    author = {Chen, Yu and Sank, D. and O'Malley, P. and White, T. and Barends, R. and Chiaro, B. and Kelly, J. and Lucero, E. and Mariantoni, M. and Megrant, A. and Neill, C. and Vainsencher, A. and Wenner, J. and Yin, Y. and Cleland, A. N. and Martinis, John M.},
    title = {Multiplexed dispersive readout of superconducting phase qubits},
    journal = {Appl. Phys. Lett.},
    volume = {101},
    number = {18},
    pages = {182601},
    year = {2012},
    month = {Nov.},
    issn = {0003-6951},
    doi = {10.1063/1.4764940},
    url = {https://doi.org/10.1063/1.4764940},
    eprint = {https://pubs.aip.org/aip/apl/article-pdf/doi/10.1063/1.4764940/14258287/182601_1_online.pdf},
}

@Article{Li2024,
author={Li, Ziqian and Roy, Tanay and Rodr{\'i}guez P{\'e}rez, David and Lee, Kan-Heng and Kapit, Eliot and Schuster, David I.},
title={Autonomous error correction of a single logical qubit using two transmons},
journal={Nat. Commun.},
year={2024},
month={Feb.},
day={23},
volume={15},
number={1},
pages={1681},
issn={2041-1723},
doi={10.1038/s41467-024-45858-z},
url={https://doi.org/10.1038/s41467-024-45858-z}
}

@Article{Rosenberg2017,
author={Rosenberg, D. and Kim, D. and Das, R. and Yost, D. and Gustavsson, S. and Hover, D. and Krantz, P. and Melville, A. and Racz, L. and Samach, G. O. and Weber, S. J. and Yan, F. and Yoder, J. L. and Kerman, A. J. and Oliver, W. D.},
title={3D integrated superconducting qubits},
journal={npj Quantum Inf.},
year={2017},
month={Oct.},
day={09},
volume={3},
number={1},
pages={42},
issn={2056-6387},
doi={10.1038/s41534-017-0044-0},
url={https://doi.org/10.1038/s41534-017-0044-0}
}

@inbook{Hong2011,
publisher = {John Wiley \& Sons, Ltd},
isbn = {9780470937297},
title = {Coupled-Resonator Circuits},
booktitle = {Microstrip Filters for RF/Microwave Applications},
author={Jia-sheng, Hong},
chapter = {7},
pages = {193-231},
year = {2011},
}

@ARTICLE{luo2025,
  author={Luo, Zhen and Richard, Léa and Tsitsilin, Ivan and Schneider, Christian M. F. and Dietz, Marco and Filipp, Stefan and Hagelauer, Amelie},
  journal={IEEE Trans. Microw. Theory Techn.}, 
  title={A Versatile Analytical Model for Fast and Accurate Determination of Feedline-Coupled Resonators for Superconducting Qubit Readout}, 
  year={2025},
  volume={73},
  number={10},
  pages={8059-8070},
  month = {Oct.},
  doi={10.1109/TMTT.2025.3578414}
}

@article{Probst2015,
    author = {Probst, S. and Song, F. B. and Bushev, P. A. and Ustinov, A. V. and Weides, M.},
    title = {Efficient and robust analysis of complex scattering data under noise in microwave resonators},
    journal = {Rev. Sci. Instrum.},
    volume = {86},
    number = {2},
    pages = {024706},
    year = {2015},
    month = {02},
    issn = {0034-6748},
    doi = {10.1063/1.4907935},
    url = {https://doi.org/10.1063/1.4907935},
    eprint = {https://pubs.aip.org/aip/rsi/article-pdf/doi/10.1063/1.4907935/15732678/024706_1_online.pdf},
}

@Article{Norris2024,
author={Norris, Graham J. and Michaud, Laurent and Pahl, David and Kerschbaum, Michael and Eichler, Christopher and Besse, Jean-Claude and Wallraff, Andreas},
title={Improved parameter targeting in 3D-integrated superconducting circuits through a polymer spacer process},
journal={EPJ Quantum Technol.},
year={2024},
month={Jan},
day={11},
volume={11},
number={1},
pages={5},
issn={2196-0763},
doi={10.1140/epjqt/s40507-023-00213-x},
url={https://doi.org/10.1140/epjqt/s40507-023-00213-x}
}

@ARTICLE{Li2023,
  author={Li, Hang-Xi and Shiri, Daryoush and Kosen, Sandoko and Rommel, Marcus and Chayanun, Lert and Nylander, Andreas and Rehammar, Robert and Tancredi, Giovanna and Caputo, Marco and Grigoras, Kestutis and Grönberg, Leif and Govenius, Joonas and Bylander, Jonas},
  journal={IEEE Trans. Quantum Eng.}, 
  title={Experimentally Verified, Fast Analytic, and Numerical Design of Superconducting Resonators in Flip-Chip Architectures}, 
  year={2023},
  volume={4},
  number={},
  pages={1-12},
  doi={10.1109/TQE.2023.3302371}}

@ARTICLE{Gevorgian1995,
  author={Gevorgian, S. and Linner, L.J.P. and Kollberg, E.L.},
  journal={IEEE Trans. Microw. Theory Techn.}, 
  title={CAD models for shielded multilayered CPW}, 
  year={1995},
  volume={43},
  number={4},
  pages={772-779},
  doi={10.1109/22.375223}}

@ARTICLE{Cheng1995,
  author={Cheng, K.-K.M. and Robertson, I.D.},
  journal={IEEE Trans. Microw. Theory Techn.}, 
  title={Quasi-TEM study of microshield lines with practical cavity sidewall profiles}, 
  year={1995},
  volume={43},
  number={12},
  month={Dec.},
  pages={2689-2694},
  doi={10.1109/22.477845}}

@ARTICLE{Cheng1997,
    author={Cheng, K.-K.M.},
    journal={IEEE Trans. Microw. Theory Techn.}, 
    title={Effect of conductor backing on the line-to-line coupling between parallel coplanar lines},
    month={Jul.},
    year={1997},
    volume={45},
    number={7},
    pages={1132-1134},
    doi={10.1109/22.598452}
}

@ARTICLE{Krupka2006,
    author={Krupka, J. and Breeze, J. and Centeno, A. and Alford, N. and Claussen, T. and Jensen, L.},
    journal={IEEE Trans. Microw. Theory Techn.}, 
    title={Measurements of Permittivity, Dielectric Loss Tangent, and Resistivity of Float-Zone Silicon at Microwave Frequencies}, 
    month={Nov.},  
    year={2006},
    volume={54},
    number={11},
    pages={3995-4001},
    doi={10.1109/TMTT.2006.883655}
}

@inproceedings{Hanna1985,
  author={Hanna, Victor Found},
  booktitle={1985 15th European Microwave Conference}, 
  title={Parameters of Coplanar Diretional Couplers with Lower Ground Plane}, 
  year={1985},
  month={Sep.},
  volume={},
  number={},
  pages={820-825},
  address={Paris, France},
  doi={10.1109/EUMA.1985.333579}}

@article{Koch2007,
  title = {Charge-insensitive qubit design derived from the Cooper pair box},
  author = {Koch, Jens and Yu, Terri M. and Gambetta, Jay and Houck, A. A. and Schuster, D. I. and Majer, J. and Blais, Alexandre and Devoret, M. H. and Girvin, S. M. and Schoelkopf, R. J.},
  journal = {Phys. Rev. A},
  volume = {76},
  issue = {4},
  pages = {042319},
  numpages = {19},
  year = {2007},
  month = {Oct.},
  publisher = {American Physical Society},
  doi = {10.1103/PhysRevA.76.042319},
  url = {https://link.aps.org/doi/10.1103/PhysRevA.76.042319}
}

@article{Walter2017,
  title = {Rapid High-Fidelity Single-Shot Dispersive Readout of Superconducting Qubits},
  author = {Walter, T. and Kurpiers, P. and Gasparinetti, S. and Magnard, P. and Poto\ifmmode \check{c}\else \v{c}\fi{}nik, A. and Salath\'e, Y. and Pechal, M. and Mondal, M. and Oppliger, M. and Eichler, C. and Wallraff, A.},
  journal = {Phys. Rev. Appl.},
  volume = {7},
  issue = {5},
  pages = {054020},
  numpages = {11},
  year = {2017},
  month = {May},
  publisher = {American Physical Society},
  doi = {10.1103/PhysRevApplied.7.054020},
  url = {https://link.aps.org/doi/10.1103/PhysRevApplied.7.054020}
}

@article{Gubin2005,
    title = {Dependence of magnetic penetration depth on the thickness of superconducting Nb thin films},
    author = {Gubin, A. I. and Il'in, K. S. and Vitusevich, S. A. and Siegel, M. and Klein, N.},
    journal = {Phys. Rev. B},
    volume = {72},
    issue = {6},
    pages = {064503},
    numpages = {8},
    year = {2005},
    month = {Aug.},
    publisher = {American Physical Society},
    doi = {10.1103/PhysRevB.72.064503},
    url = {https://link.aps.org/doi/10.1103/PhysRevB.72.064503}
}

@article{Pinto_2018,
    author = {Pinto, Nicola and Rezvani, S. J and Perali, Andrea and Flammia, Luca and Milošević, Milorad and Fretto, M. and Cassiago, Cristina and Leo, Natascia},
    year = {2018},
    month = {Mar.},
    pages = {},
    number = {1},
    title = {Dimensional crossover and incipient quantum size effects in superconducting niobium nanofilms},
    volume = {8},
    journal = {Scientific Reports},
    doi = {10.1038/s41598-018-22983-6}
}

@article{Keiji1992,
    doi = {10.1143/JJAP.31.3844},
    url = {https://dx.doi.org/10.1143/JJAP.31.3844},
    year = {1992},
    month = {Dec.},
    publisher = {},
    volume = {31},
    number = {12R},
    pages = {3844},
    author = {Yoshida, Keiji and Hossain, Mohammad Sajjad and Kisu, Takanobu and Keiji Enpuku, Keiji Enpuku and Kaoru Yamafuji, Kaoru Yamafuji},
    title = {Modeling of Kinetic-Inductance Coplanar Striplin with NbN Thin Films},
    journal = {Jpn. J. Appl. Phys.},
}

@article{Huang2021,
  title = {Microwave Package Design for Superconducting Quantum Processors},
  author = {Huang, Sihao and Lienhard, Benjamin and Calusine, Greg and Veps\"al\"ainen, Antti and Braum\"uller, Jochen and Kim, David K. and Melville, Alexander J. and Niedzielski, Bethany M. and Yoder, Jonilyn L. and Kannan, Bharath and Orlando, Terry P. and Gustavsson, Simon and Oliver, William D.},
  journal = {PRX Quantum},
  volume = {2},
  issue = {2},
  pages = {020306},
  numpages = {16},
  year = {2021},
  month = {Apr},
  publisher = {American Physical Society},
  doi = {10.1103/PRXQuantum.2.020306},
  url = {https://link.aps.org/doi/10.1103/PRXQuantum.2.020306}
}

@INPROCEEDINGS{Benjamin2019,
  author={Lienhard, Benjamin and Braumüller, Jochen and Woods, Wayne and Rosenberg, Danna and Calusine, Greg and Weber, Steven and Vepsäläinen, Antti and O'Brien, Kevin and Orlando, Terry P. and Gustavsson, Simon and Oliver, William D.},
  booktitle={2019 IEEE MTT-S International Microwave Symposium (IMS)}, 
  title={Microwave Packaging for Superconducting Qubits},
  address = {Boston, MA, USA},
  year={2019},
  volume={},
  number={},
  pages={275-278},
  doi={10.1109/MWSYM.2019.8701119}}

@article{Goppl2008,
    author = {Göppl, M. and Fragner, A. and Baur, M. and Bianchetti, R. and Filipp, S. and Fink, J. M. and Leek, P. J. and Puebla, G. and Steffen, L. and Wallraff, A.},
    title = {Coplanar waveguide resonators for circuit quantum electrodynamics},
    journal = {J. Appl. Phys.},
    volume = {104},
    number = {11},
    pages = {113904},
    year = {2008},
    month = {Dec.},
    issn = {0021-8979},
    doi = {10.1063/1.3010859},
    url = {https://doi.org/10.1063/1.3010859},
}

@phdthesis{Gao2008,
  author  = {Jiansong Gao},
  title   = {The Physics of Superconducting Microwave Resonators},
  school  = {Dept. Phys., California Institute of Technology},
  address = {Pasadena, CA, USA},
  year    = {2008},
}

@article{Shima2025,
    author = {Poorgholam-Khanjari, Shima and Seferai, Valentino and Foshat, Paniz and Rose, Calum and Feng, Hua and Hadfield, Robert H. and Weides, Martin and Delfanazari, Kaveh},
    title = {Engineering high-Q superconducting tantalum microwave coplanar waveguide resonators for compact coherent quantum circuit},
    journal = {Sci. Rep.},
    year = {2025},
    volume = {15},
    number = {1},
    pages = {27113},
    doi = {10.1038/s41598-025-11744-x},
    url = {https://doi.org/10.1038/s41598-025-11744-x},
    issn = {2045-2322},
}

@ARTICLE{Amini2022,
    author={Amini, Mohammad Hossein and Davoodi-Rad, Zahra and Ghasemi, Amirhossein and Khaliliy, Mohsen and Mallahzadeh, Alireza},
    journal={IEEE Trans. Appl. Supercond.}, 
    title={Current Distribution and Kinetic Inductance of Coplanar Waveguide Transmission Lines}, 
    year={2022},
    month = {Nov.},
    volume={32},
    number={8},
    pages={1-6},
    doi={10.1109/TASC.2022.3203029}
}

@article{Wenner2011,
    doi = {10.1088/0953-2048/24/6/065001},
    url = {https://doi.org/10.1088/0953-2048/24/6/065001},
    year = {2011},
    month = {Mar.},
    publisher = {},
    volume = {24},
    number = {6},
    pages = {065001},
    author = {Wenner, J and Neeley, M and Bialczak, Radoslaw C and Lenander, M and Lucero, Erik and O’Connell, A D and Sank, D and Wang, H and Weides, M and Cleland, A N and Martinis, John M},
    title = {Wirebond crosstalk and cavity modes in large chip mounts for superconducting
    qubits},
    journal = {Supercond. Sci. Technol.},
}

\vfill

\end{document}